\documentclass[12pt]{article}
\usepackage{amsfonts}
\usepackage{amsmath}
\usepackage{amssymb}
\usepackage{graphicx}
\usepackage{color}
\usepackage[all, knot]{xy}
\usepackage{tikz}

\usepackage{ulem}

\usepackage[utf8]{inputenc}
\usepackage{epstopdf}
\usepackage[footnotesize]{caption}
\usepackage{amsthm}
\usepackage{enumitem}
\usepackage{mathrsfs}

\usepackage[margin=3cm]{geometry}

\def \be {\begin{equation}}
\def \ee {\end{equation}}
\def \bea {\begin{eqnarray}}
\def \eea {\end{eqnarray}}
\def \nn {\nonumber}

\def \rr {\raise.35ex\hbox{\small $\prime$}\kern-.17em{\mbox{\large $\imath$}}}

\def \dels {\partial\kern-.6em /\kern.1em}
\def \As {{A\kern-.5em / \kern.5em}}
\def \Ds {D\kern-.7em / \kern.5em}

\def \ks {k\kern-.5em /}
\def \ls {l\kern-.5em /}

\newcommand{\ci}[1]{}

\newcommand{\ba}{\begin{eqnarray}}
\newcommand{\ea}{\end{eqnarray}}
\newcommand{\bal}{\begin{align}}
\newcommand{\eal}{\end{align}}
\newcommand{\bay}[1]{\left(\begin{array}{#1}}
\newcommand{\eay}{\end{array}\right)}

\newcommand{\hide}[1]{}

\DeclareMathOperator{\sech}{sech}

\newlist{axioms}{enumerate}{2}
\setlist[axioms,1]{label=\textbf{A\arabic{axiomsi}.}, ref=A\arabic{axiomsi}}
\setlist[axioms,2]{label=\textbf{A\arabic{axiomsi}\rlap{\myEnumCounter{axiomsii}}.},%
                   ref=A\arabic{axiomsi}\myEnumCounter{axiomsii},%
                   align=parleft,%
                   leftmargin=0em,%
                   itemsep=1.4ex,%
                   before={\stepcounter{axiomsi}}}

  \usetikzlibrary{decorations.markings}

\begin{document}

\begin{titlepage}
\begin{center}

\textbf{\LARGE
U(1) CS Theory vs SL(2) CS Formulation: Boundary Theory and Wilson Line
\vskip.3cm
}
\vskip .5in
{\large
Xing Huang$^{a, b, c}$ \footnote{e-mail address: xingavatar@gmail.com},
Chen-Te Ma$^{d, e, f, g, h}$ \footnote{e-mail address: yefgst@gmail.com}, 
Hongfei Shu$^{i, j, k, l}$ \footnote{e-mail address: hongfei.shu@su.se}, and
\\ Chih-Hung Wu$^{m}$ \footnote{e-mail address: chih-hungwu@physics.ucsb.edu}
\\
\vskip 1mm
}
{\sl
$^a$
Institute of Modern Physics, Northwest University, Xi'an 710069, China.
\\
$^b$
Shaanxi Key Laboratory for Theoretical Physics Frontiers, Xi'an 710069, China. 
\\
$^c$
NSFC-SPTP Peng Huanwu Center for Fundamental Theory, Xi'an 710127, China.
\\
$^d$
Asia Pacific Center for Theoretical Physics,\\
Pohang University of Science and Technology, 
Pohang 37673, Gyeongsangbuk-do, South Korea. 
\\
$^e$
Guangdong Provincial Key Laboratory of Nuclear Science,\\
 Institute of Quantum Matter,
South China Normal University, Guangzhou 510006, Guangdong, China.
\\
$^f$
School of Physics and Telecommunication Engineering,\\
 South China Normal University, Guangzhou 510006, Guangdong, China.
\\
$^g$
Guangdong-Hong Kong Joint Laboratory of Quantum Matter,\\
 Southern Nuclear Science Computing Center, 
South China Normal University, Guangzhou 510006, China.
\\
$^h$
The Laboratory for Quantum Gravity and Strings,\\
 Department of Mathematics and Applied Mathematics,\\
University of Cape Town, Private Bag, Rondebosch 7700, South Africa.
\\
$^i$
Beijing Institute of Mathematical Sciences and Applications (BIMSA), Beijing, 101408, China. 
\\
$^j$
Yau Mathematical Sciences Center (YMSC), Tsinghua University, Beijing, 100084, China.
\\
$^k$
Nordita, KTH Royal Institute of Technology and Stockholm University,\\
 Roslagstullsbacken 23, SE-106 91 Stockholm, Sweden.
\\
$^l$
Department of Physics, Tokyo Institute of Technology, Tokyo, 152-8551, Japan.
\\
$^m$
Department of Physics, University of California, Santa Barbara, CA 93106, USA.
}
\\
\vskip 1mm
\vspace{40pt}
\end{center}
\newpage
\begin{abstract}
We first derive the boundary theory from the U(1) Chern-Simons theory. 
The boundary action on an $n$-sheet manifold appears from its back-reaction of the Wilson line. 
The reason is that the U(1) Chern-Simons theory can provide an exact effective action when introducing the Wilson line. 
The Wilson line in the pure AdS$_3$ Einstein gravity is equivalent to entanglement entropy in the boundary theory up to classical gravity.
The U(1) Chern-Simons theory deviates by a self-interaction term from the gauge formulation on the boundary.   
We also compare the Hayward term in the SL(2) Chern-Simons formulation to the Wilson line approach. 
Introducing two wedges can reproduce the entanglement entropy for a single interval at the classical level. 
We propose quantum generalization by combining the bulk and Hayward terms. 
The quantum correction of the partition function vanishes.    
In the end, we calculate the entanglement entropy for a single interval. 
The pure AdS$_3$ Einstein gravity theory shows a shift of central charge by 26 at the one-loop level. 
The U(1) Chern-Simons theory does not show a shift from the quantum effect. 
The result is the same in the weak gravitational constant limit. 
The non-vanishing quantum correction shows that the Hayward term is incorrect.
\end{abstract}
\end{titlepage}

\section{Introduction}
\label{sec:1}
\noindent
The goal of studying {\it emergent spacetime} is to obtain the bulk geometry from other equivalent descriptions.
Since Einstein gravity theory is not renormalizable, people expect that Einstein gravity theory may not be a fundamental theory. 
One promising approach is the {\it Holographic Principle} \cite{Bekenstein:1973ur,Bardeen:1973gs,Hawking:1974sw,tHooft:1993dmi,Susskind:1994vu}. 
This principle states that the physical degrees of freedom in Quantum Gravity are all on its boundary.
One candidate for perturbative Quantum Gravity, String Theory, provided a conjecture, the {\it Anti-de Sitter/Conformal Field Theory Correspondence} (AdS/CFT correspondence). 
String theory in the $(d+1)$-dimensional AdS manifold is dual to CFT$_{d}$. 
It is a surprising fact that a gravitational theory is calculable. 
Furthermore, the Ryu-Takayanagi prescription shows that the surface with a minimum area (minimum surface) in AdS$_{d+1}$ contains the boundary information (entanglement entropy) \cite{Ryu:2006bv,Ryu:2006ef}. 
Therefore, this seems to imply {\it Quantum Entanglement} generates the bulk spacetime. 
\\

\noindent
The calculation of quantum entanglement quantities, in general, relies on the {\it $n$-sheet manifold} with the analytical continuation of $n$ \cite{Holzhey:1994we}, which is {\it hard} to calculate. 
The minimum surface provides a simplified way to study entanglement entropy \cite{Lewkowycz:2013nqa,Casini:2011kv}. 
However, to confirm the observation, it is necessary to perform a perturbation in a boundary theory to probe the quantum regime of bulk gravity theory. 
To obtain {\it exact properties} in Quantum Entanglement, we study Chern-Simons (CS) Theory \cite{Elitzur:1989nr}.
\\

\noindent
We begin with the AdS$_3$ Einstein gravity theory. 
Because 3d pure Einstein gravity theory (metric formulation) does not have local degrees of freedom \cite{Brown:1986nw}. 
The metric formulation is equivalent to an SL(2) CS formulation (gauge formulation) up to the classical level \cite{Witten:1988hc}. 
Although the metric formulation is not equivalent to the gauge formulation due to its measure, the renormalizable gauge formulation is convenient for computation in the {\it quantum regime} \cite{Giombi:2008vd, Huang:2019nfm}. 
\\

\noindent
The absence of local degrees of freedom implies that one can derive the boundary action.
The first derivation showed that the boundary theory is the Liouville theory \cite{Coussaert:1995zp}, CFT. 
Because the Liouville theory does not have a normalizable vacuum, it contradicts the bulk theory \cite{Witten:2007kt}. 
Indeed, it is a classical action because people do not consider the measure. 
Recently, due to a proper consideration for including the measure into the boundary theory, the {\it 2d Schwarzian theory} appears. 
This theory has the expected {\it SL(2) reparametrization} gauge redundancy \cite{Cotler:2018zff}. 
The 2d Schwarzian theory does not have {\it conformal symmetry}. 
People should expect the result due to conformal anomaly. 
\\

\noindent 
The first question is to find the {\it gauge version of the minimum surface}. 
Because the {\it Wilson lines} constitute all operators in the gauge formulation, {\it Wilson line} is shown to be a probe of holographic entanglement \cite{Ammon:2013hba,deBoer:2013vca}. 
In the AdS/CFT correspondence, it is not a simple task to study the minimum surfaces at an {\it operator level}. 
The CS theory is a simple system for avoiding difficulty.
Hence studying the gauge formulation of pure AdS$_3$ Einstein gravity theory should give a clue to emergent spacetime.  
\\

\noindent
The co-dimension two Hayward term \cite{Hayward:1993my} was recently connected with the entanglement entropy at the classical gravity level \cite{Takayanagi:2019tvn}. 
The basic idea is to introduce a Euclidean bulk manifold with two boundaries that are not smoothly junctioned. 
When gluing two boundary surfaces at a co-dimension two wedge, the junction conditions induce an additional Hayward term. It ensures a well-posed variational principle. 
The Hayward term induces the wedge contribution to the partition function \cite{Takayanagi:2019tvn,Botta-Cantcheff:2020ywu}. 
This contribution implies the equivalence with the co-dimension two cosmic brane action with a suitable tension \cite{Takayanagi:2019tvn,Botta-Cantcheff:2020ywu}. 
The approach effectively introduces the back-reaction on the replica orbifold that generates a conical geometry \cite{Dong:2016fnf}. 
The Hayward term is a gravitational action rather than an auxiliary tool (cosmic brane action). 
Therefore, we are interested in understanding how the wedge term arises in the SL(2) CS formulation. 
\\

\noindent
In this paper, we study holographic entanglement entropy in the gauge formulation. 
We calculate entanglement entropy and Wilson line with the quantum correction. 
The SL(2) CS formulation has exact solutions to the boundary effective action and the partition function \cite{Cotler:2018zff}. 
However, it does not have an exact solution when introducing the Wilson lines. 
Introducing the Wilson line produces a back-reaction to generate an $n$-sheet manifold \cite{Ammon:2013hba}. 
The additional term related to the Wilson line vanishes in the limit $n\rightarrow 1$. 
Hence this does not produce a problem but relies on whether the limit or a smooth solution exists. 
We will first study a similar but simpler system, U(1) CS theory.  
Furthermore, to understand whether the Hayward term is a suitable proposal for obtaining entanglement entropy, we study the Hayward term in the SL(2) CS formulation.
To summarize our results:
\begin{itemize} 
\item We derive the boundary action of U(1) CS theory. 
The result shows a combination of left- and right-chiral non-interacting scalar field theories \cite{Floreanini:1987as}. 
This theory can be rewritten from the non-interacting scalar field theory \cite{Tseytlin:1990ar}. 
Introducing the Wilson lines is equivalent to deforming the chiral scalar fields and does not change the bulk theory. 
Obtaining the non-interacting scalar field theory on an $n$-sheet manifold is necessary to deform the boundary condition. 
We perform this study by first deforming the chiral scalar fields and choosing the $n$-sheet background. 
It establishes an exact correspondence between the Wilson line and entanglement entropy. 
This study shows that deformation is crucial for the relation between the Wilson line and entanglement entropy.
\item We derive the boundary action for the SL(2) CS formulation. 
The boundary action is similar to the U(1) case (but it has additional terms for the interaction). 
We can choose the weak gravitational constant limit to truncate the self-interaction. 
The non-trivial deformation is again necessary for building an exact correspondence between the Wilson line and entanglement entropy as in the U(1) CS theory. 
Indeed, this establishes the ``minimum surface=entanglement entropy'' prescription with the quantum fluctuation of gauge fields for studying the quantum deformation of minimum surface \cite{Huang:2019nfm}. 
\item We use the boundary term \cite{Rooman:2000zi} to obtain the co-dimension two Hayward term. 
This method is the same as in the metric formulation. 
Two wedges forming a single interval on the boundary can reproduce the entanglement entropy at the classical level. 
Combining the bulk and Hayward terms does not generate quantum correction. 
It is inconsistent with the one-loop exact calculation of the entanglement entropy. 
Hence we conclude that the quantum generalization does not provide a suitable description of the entanglement entropy.
\item We calculate the entanglement entropy for a single interval in the SL(2) CS formulation. 
It shows an exact shift of central charge by 26. 
The shift should result from the additional interacting term since the U(1) CS theory does not generate the shifting. 
This result supports the equivalence of the two theories in the weak gravitational constant limit. 
The U(1) CS theory and the SL(2) CS formulation do not have the same number on gauge fields. 
The boundary degrees of freedom are the same. 
Hence showing equivalence is a non-trivial result.    
\end{itemize}

\noindent
The organization of this paper is as follows: We derive the boundary action and study its dual in Sec.~\ref{sec:2}. 
The duality provides the relation between the Wilson line and entanglement entropy.  
We then consider the boundary action, the building of ``minimum surface=entanglement entropy'' for SL(2) CS formulation, and the set-up of the co-dimension two Hayward term in Sec.~\ref{sec:3}. 
We calculate the entanglement entropy for a single interval in SL(2) CS formulation at one loop in Sec.~\ref{sec:4}. 
In the end, we have our concluding remarks in Sec.~\ref{sec:5}. 
We give the details of the one-loop correction for Rényi entropy in Appendix \ref{appa}.

\section{U(1) CS Theory}
\label{sec:2}
\noindent
To provide a clear picture of the bulk-boundary correspondence, we first introduce U(1) CS theory. 
We derive the boundary action in U(1) CS theory. 
We then introduce the Wilson lines for a study of entanglement entropy.

\subsection{Action}
\noindent
The action for the U(1) CS theory in the Lorentzian manifold is given by
\bea
&&S_{\mathrm{U}(1)}
\nn\\
&=&\frac{k}{2\pi}\int d^3x\ \bigg(A_tF_{r\theta}-\frac{1}{2}\big(A_r\partial_tA_{\theta}-A_{\theta}\partial_tA_r\big)\bigg)
\nn\\
&&-\frac{k}{2\pi}\int d^3x\ \bigg(\bar{A}_t\bar{F}_{r\theta}-\frac{1}{2}\big(\bar{A}_r\partial_t\bar{A}_{\theta}-\bar{A}_{\theta}\partial_t\bar{A}_r\big)\bigg)
\nn\\
&&-\frac{k}{4\pi}\int dtd\theta\ \frac{E_t^+}{E_{\theta}^+}A_{\theta}^2
\nn\\
&&+\frac{k}{4\pi}\int dtd\theta\ \frac{E_t^-}{E_{\theta}^-}\bar{A}_{\theta}^2.
\label{SU1}
\eea
The boundary conditions become:
\bea
(E_{\theta}^+A_t-E_t^+A_{\theta})|_{r\rightarrow\infty}=0; \qquad (E_{\theta}^-\bar{A}_t-E_{t}^-\bar{A}_{\theta})|_{r\rightarrow\infty}=0.
\eea
The $t$, $r$, $\theta$ are the polar coordinates, and the ranges are given by:
\bea
-\infty<t<\infty; \qquad 0<r<\infty; \qquad 0<\theta\le 2\pi.
\eea
The boundary zweibein is defined by:
\bea
E^+\equiv E^{\theta}+E^t; \qquad E^{-}\equiv E^{\theta}-E^t.
\eea
We also define the component of the boundary zweibein as the following:
\bea
E^+_{\theta}d\theta\equiv E^{\theta}; \qquad 
E^+_t dt\equiv E^t; \qquad 
E^-_{\theta} d\theta\equiv E^{\theta}; \qquad 
E^-_t dt\equiv -E^t.
\eea
The boundary metric in terms of the boundary zweibein is
\bea
g_{\tilde{\mu}\tilde{\nu}}\equiv \frac{1}{2}(E_{\tilde{\mu}}^+E_{\tilde{\nu}}^-+E_{\tilde{\mu}}^-E_{\tilde{\nu}}^+).
\eea
The indices of boundary spacetime are $\tilde{\mu}=t, \theta$. 
The notation for the $x^+$ and $x^-$ is:
\bea
x^+=t+\theta; \qquad x^-=t-\theta.
\eea 
We can do the integration by part
\bea
\frac{k}{2\pi}\int d^3x\ A_t F_{r\theta}=\frac{k}{4\pi}\int d^3x\ \bigg(A_t\partial_r A_{\theta}-A_t\partial_{\theta}A_r
-A_{\theta}\partial_rA_t+A_r\partial_{\theta}A_t\bigg)
\eea
 to show the familiar CS form
 \bea
 S_{\mathrm{U(1)}}=\frac{k}{4\pi}\int d^3x\ \epsilon^{\mu\nu\rho}A_{\mu}\partial_{\nu}A_{\rho}, 
 \eea
 where $\mu, \nu, \rho$ are the indices for the bulk coordinates.

\subsection{Bulk-Boundary Correspondence}
\noindent
Integrating out $A_t$ and $\bar{A}_t$ respectively, we obtain:
\bea
F_{r\theta}=0; \qquad \bar{F}_{r\theta}=0.
\eea
The solutions are: 
\bea
A_r=g\partial_rg^{-1}; \qquad A_{\theta}=g\partial_{\theta}g^{-1}; \qquad \bar{A}_r=-\bar{g}\partial_r\bar{g}^{-1}; \qquad \bar{A}_{\theta}=-\bar{g}\partial_{\theta}\bar{g}^{-1},
\eea
where $g$ and $\bar{g}$ are reparametrized by $\phi$ and $\bar{\phi}$ respectively:
\bea
g\equiv e^{-\phi}; \qquad \bar{g}\equiv e^{\bar{\phi}}.
\eea
Hence the solutions become:
\bea
A_r=\partial_r\phi; \qquad A_{\theta}=\partial_{\theta}\phi; \qquad \bar{A}_r=\partial_r\bar{\phi}; \qquad \bar{A}_{\theta}=\partial_{\theta}\bar{\phi}.
\eea
By substituting the solutions into Eq. \eqref{SU1}, the action becomes:
\bea
&&S_{\mathrm{U}(1)}
\nn\\
&=&-\frac{k}{4\pi}\int d^3x\ (\partial_r\phi\partial_t\partial_{\theta}\phi-\partial_{\theta}\phi\partial_t\partial_r\phi)
+\frac{k}{4\pi}\int d^3x\ (\partial_r\bar{\phi}\partial_t\partial_{\theta}\bar{\phi}-\partial_{\theta}\bar{\phi}\partial_t\partial_r\bar{\phi})
\nn\\
&&-\frac{k}{4\pi}\int dtd\theta\ \frac{E_t^+}{E_{\theta}^+}\partial_{\theta}\phi\partial_{\theta}\phi 
+\frac{k}{4\pi}\int dtd\theta\ \frac{E_t^-}{E_{\theta}^-}\partial_{\theta}\bar{\phi}\partial_{\theta}\bar{\phi}
\nn\\
&=&-\frac{k}{4\pi}\int d^3x\ (-\partial_{\theta}\partial_r\phi\partial_t\phi-\partial_{\theta}\phi\partial_t\partial_r\phi)
\nn\\
&&
+\frac{k}{4\pi}\int d^3x\ (-\partial_{\theta}\partial_r\bar{\phi}\partial_t\bar{\phi}-\partial_{\theta}\bar{\phi}\partial_t\partial_r\bar{\phi})
\nn\\
&&-\frac{k}{4\pi}\int dtd\theta\ \frac{E_t^+}{E_{\theta}^+}\partial_{\theta}\phi\partial_{\theta}\phi 
+\frac{k}{4\pi}\int dtd\theta\ \frac{E_t^-}{E_{\theta}^-}\partial_{\theta}\bar{\phi}\partial_{\theta}\bar{\phi}
\nn\\
&=&
\frac{k}{4\pi}\int dtd\theta\ \partial_{\theta}\phi\partial_t\phi
-\frac{k}{4\pi}\int dtd\theta\ \partial_{\theta}\bar{\phi}\partial_t\bar{\phi}
\nn\\
&&
-\frac{k}{4\pi}\int dtd\theta\ \frac{E_t^+}{E_{\theta}^+}\partial_{\theta}\phi\partial_{\theta}\phi 
+\frac{k}{4\pi}\int dtd\theta\ \frac{E_t^-}{E_{\theta}^-}\partial_{\theta}\bar{\phi}\partial_{\theta}\bar{\phi}
\nn\\
&=&\frac{k}{4\pi}\int dtd\theta\ \partial_{\theta}\phi\bigg(\partial_t-\frac{E_t^+}{E_{\theta}^+}\partial_{\theta}\bigg)\phi
-\frac{k}{4\pi}\int dtd\theta\ \partial_{\theta}\bar{\phi}\bigg(\partial_t-\frac{E_t^-}{E_{\theta}^-}\partial_{\theta}\bigg)\bar{\phi}.
\eea
In the second equality, we have performed an integration by parts. 
Integrating by part generates a boundary term $\sim\partial_r\phi\partial_t\phi$. 
For the compact U(1), the $\phi$ has a constant shifting. 
The boundary term does not have a contribution in general. 
Hence the effective action combines the left- and right-chiral scalar field theories. 
In the SL(2) CS formulation, we choose the asymptotic gauge field for obtaining the AdS$_3$ metric. 
It is the main point for obtaining the equivalence in boundary degrees of freedom between the U(1) CS theory and SL(2) CS formulation.
We will discuss the SL(2) CS formulation later.

\subsection{Non-Chiral Scalar Field Theory and $n$-Sheet Manifold}
\noindent
The $n$-sheet cylindrical manifold is:
\bea
ds^2=-dt^2+n^2d\theta^2, \ 
E_t^+E_t^-=-1, \ 
E_{\theta}^+E_{\theta}^-=n^2.
\eea
We can redefine the 
\bea
\tilde{\theta}\equiv n\theta,
\eea
and then the metric becomes the flat background, but the range of the angular coordinate changes.
The non-interacting scalar field theory with a curved background is:
\bea
S_{\mathrm{U}(1)c}
&=&-\frac{k}{8\pi}\int dtd\theta\sqrt{-\det g_{\tilde{\mu}\tilde{\nu}}}\ g^{\tilde{\mu}\tilde{\nu}}
\partial_{\tilde{\mu}}\Phi
\partial_{\tilde{\nu}}\Phi
\nn\\
&=&
-\frac{nk}{8\pi}\int dtd\theta\ 
\bigg(-\partial_t\Phi\partial_t\Phi
+\frac{1}{n^2}\partial_{\theta}\Phi\partial_{\theta}\Phi\bigg),
\eea
obtained by integrating out the auxiliary field $\Pi$ from the following action
 \bea
 S_{\mathrm{U}(1)c1}=
 \frac{nk}{4\pi}\int dtd\theta\ 
 \bigg(\Pi\partial_t\Phi-\frac{1}{2}\Pi^2
 -\frac{1}{2n^2}\partial_{\theta}\Phi\partial_{\theta}\Phi\bigg).
 \eea
 Now we apply the field redefinition:
 \bea
 \Phi\equiv\phi+\bar{\phi}; \qquad 
 \Pi\equiv\frac{1}{n}\partial_{\theta}(\phi-\bar{\phi})
 \eea
 to obtain:
 \bea
&&\Pi\partial_t\Phi-\frac{1}{2}\Pi^2-\frac{1}{2n^2}\partial_{\theta}\Phi\partial_{\theta}\Phi
\nn\\
&=&\frac{1}{n}(\partial_{\theta}\phi-\partial_{\theta}\bar{\phi})(\partial_t\phi+\partial_t\bar{\phi})
\nn\\
&&
-\frac{1}{2n^2}(\partial_{\theta}\phi\partial_{\theta}\phi
+\partial_{\theta}\bar{\phi}\partial_{\theta}\bar{\phi}
-2\partial_{\theta}\phi\partial_{\theta}\bar{\phi})
\nn\\
&&
-\frac{1}{2n^2}(\partial_{\theta}\phi\partial_{\theta}\phi
+\partial_{\theta}\bar{\phi}\partial_{\theta}\bar{\phi}
+2\partial_{\theta}\phi\partial_{\theta}\bar{\phi})
\nn\\
&=&
\frac{1}{n}\partial_{\theta}\phi\bigg(\partial_t-\frac{1}{n}\partial_{\theta}\bigg)\phi
-\frac{1}{n}\partial_{\theta}\bar{\phi}\bigg(\partial_t+\frac{1}{n}\partial_{\theta}\bigg)\bar{\phi}
+\frac{1}{n}\partial_{\theta}\phi\partial_t\bar{\phi}
-\frac{1}{n}\partial_t\phi\partial_{\theta}\bar{\phi};
\nn
\eea
then
\bea
&&\frac{nk}{4\pi}\int dtd\theta\ 
\bigg(\Pi\partial_t\Phi-\frac{1}{2}\Pi^2-\frac{1}{2n^2}\partial_{\theta}\Phi\partial_{\theta}\Phi\bigg)
\nn\\
&=&\frac{k}{4\pi}\int dtd\theta\ \partial_{\theta}\phi\bigg(\partial_t-\frac{1}{n}\partial_{\theta}\bigg)\phi
-\frac{k}{4\pi}\int dtd\theta\ \partial_{\theta}\bar{\phi}\bigg(\partial_t+\frac{1}{n}\partial_{\theta}\bigg)\bar{\phi}
\nn\\
&&
+\frac{k}{4\pi}\int dt d\theta\ (\partial_{\theta}\phi\partial_t\bar{\phi}
-\partial_t\phi\partial_{\theta}\bar{\phi})
\nn\\
&=&
\frac{k}{4\pi}\int dtd\theta\ \partial_{\theta}\phi\bigg(\partial_t-\frac{1}{n}\partial_{\theta}\bigg)\phi
-\frac{k}{4\pi}\int dtd\theta\ \partial_{\theta}\bar{\phi}\bigg(\partial_t+\frac{1}{n}\partial_{\theta}\bigg)\bar{\phi}.
\eea
The last equality is up to a total derivative term. 
By comparing the action with Eq. \eqref{SU1}, we obtain the action $S_{\mathrm{U}(1)}$ with the following choices of background:
\bea
\frac{E_t^+}{E_{\theta}^+}=\frac{1}{n}; \qquad \frac{E_t^-}{E_{\theta}^-}=-\frac{1}{n}.
\eea
Therefore, we can choose the following solution:
\bea
E_t^+=1; \qquad E_{\theta}^+=n; \qquad E_t^-=-1; \qquad E_{\theta}^-=n.
\eea
 
\subsection{Wilson Lines}
\noindent
We calculate the expectation value of the Wilson line in the U(1) CS theory
\bea
\langle W\rangle\equiv\frac{1}{Z_{\mathrm{U}(1)}}\exp\big(i S_{\mathrm{U}(1)}+\sqrt{2c_2}\ln W(P, Q)\big),
\eea
where $Z_{\mathrm{U}(1)}$ is the partition function of the U(1) CS theory, and
\bea
\sqrt{2c_2}\equiv \frac{c}{6}(1-n),
\eea
in which $c$ is just a real-valued number,
and
\bea
W(P, Q)\equiv{\cal P}\exp\bigg(\int^{s(P)}_{s(Q)}ds\ \frac{dx^{\mu}}{ds}\bar{A}_{\mu}\bigg)
\exp\bigg(\int_{s(P)}^{s(Q)}ds\ \frac{dx^{\nu}}{ds}\ A_{\nu}\bigg),
\eea
where the ${\cal P}$ denotes a path-ordering, and $P, Q$ are the two end points of the Wilson lines on a time slice.
The equations of motion are:
\bea
\frac{k}{2\pi}F_{\nu\rho}&=&i\sqrt{2c_2}\epsilon_{\mu\nu\rho}\int_{s(P)}^{s(Q)}ds\ \frac{dx^{\mu}}{ds}\delta\big(x-x(s)\big); 
\nn\\ 
\frac{k}{2\pi}\bar{F}_{\nu\rho}&=&-i\sqrt{2c_2}\epsilon_{\mu\nu\rho}\int_{s(Q)}^{s(P)}ds\ \frac{dx^{\mu}}{ds}\delta\big(x-x(s)\big).
\eea
The solution is: 
\bea
A=gag^{-1}+gdg^{-1}; \qquad \bar{A}=-\bar{g}a\bar{g}^{-1}-\bar{g}d\bar{g}^{-1},
\eea
where
\bea
&&
g=e^{\phi};\qquad
\bar{g}=e^{\bar{\phi}};\qquad 
a\equiv\sqrt{\frac{c_2}{2}}\frac{1}{k}\bigg(\frac{dz}{z}-\frac{d\bar{z}}{\bar{z}}\bigg); 
\nn\\
&&
z\equiv\theta+i\psi;\qquad \bar{z}\equiv\theta-i\psi.
\nn\\
\eea
The Euclidean time is 
\bea
\psi\equiv it.
\eea
Hence we rewrite the solution like the following:
\bea
A=a+d\phi; \qquad \bar{A}=-a+d\bar{\phi},
\eea
in which the gauge field $a$ has the consistent holonomy
\bea
\int a=2\pi i\frac{\sqrt{2c_2}}{k}.
\eea
Writing the components of the gauge fields, the solution is:
\bea
A_r&=&\partial_r\phi,
\nn\\
A_{\theta}&=&\partial_{\theta}\phi+\sqrt{\frac{c_2}{2}}\frac{1}{k}\bigg(\frac{1}{\theta+i\psi}-\frac{1}{\theta-i\psi}\bigg),
\nn\\
A_{\psi}&=&\partial_{\psi}\phi+i\sqrt{\frac{c_2}{2}}\frac{1}{k}\bigg(\frac{1}{\theta+i\psi}+\frac{1}{\theta-i\psi}\bigg);
\nn\\
\bar{A}_r&=&\partial_r\bar{\phi},
\nn\\
\bar{A}_{\theta}&=&\partial_{\theta}\bar{\phi}-\sqrt{\frac{c_2}{2}}\frac{1}{k}\bigg(\frac{1}{\theta+i\psi}-\frac{1}{\theta-i\psi}\bigg),
\nn\\
\bar{A}_{\psi}&=&\partial_{\psi}\bar{\phi}-i\sqrt{\frac{c_2}{2}}\frac{1}{k}\bigg(\frac{1}{\theta+i\psi}+\frac{1}{\theta-i\psi}\bigg).
\eea
We can introduce the new scalar fields as:
\bea
\tilde{\phi}=\phi+\sqrt{\frac{c_2}{2}}\frac{1}{k}\ln\frac{\theta+i\psi}{\theta-i\psi}; \qquad 
\bar{\tilde{\phi}}=\bar{\phi}-\sqrt{\frac{c_2}{2}}\frac{1}{k}\ln\frac{\theta+i\psi}{\theta-i\psi}.
\eea
We rewrite the gauge fields in terms of the new scalar fields:
\bea
&&
A_r=\partial_r\tilde{\phi}, \qquad
A_{\theta}=\partial_{\theta}\tilde{\phi}, \qquad
A_{\psi}=\partial_{\psi}\tilde{\phi};
\nn\\
&&
\bar{A}_r=\partial_r\bar{\tilde{\phi}}, \qquad
\bar{A}_{\theta}=\partial_{\theta}\bar{\tilde{\phi}}, \qquad
\bar{A}_{\psi}=\partial_{\psi}\bar{\tilde{\phi}}.
\eea
Therefore, the Wilson line gives a back-reaction or a redefinition of the fields. 
The non-chiral scalar field theory on an $n$-sheet cylindrical manifold does not change the form of the action. 
Therefore, we also need to do the same deformation of the boundary conditions as the following:
\bea
A_t=iA_{\psi}=A_{\tilde{\theta}}; \qquad \bar{A}_t=i\bar{A}_{\psi}=-\bar{A}_{\tilde{\theta}}.
\eea
We have chosen the $n$-sheet cylindrical background in the above boundary condition. 
The deformation is crucial for establishing the relation between the Wilson line and entanglement entropy.
\\

\noindent
We want to discuss the relation between geometry and the chiral scalar fields.
The metric in the SL(2) CS formulation is \cite{Witten:1988hc}:
\bea
g_{\tilde{\mu}\tilde{\nu}}\equiv 2e_{\tilde{\mu}}e_{\tilde{\nu}}; \qquad 
A_{\mu}\equiv e+\omega; \qquad 
\bar{A}_{\mu}\equiv e-\omega.
\eea
The $e_{\tilde{\mu}}$ is vielbein, and $\omega_{\tilde{\mu}}$ is the spin connection.  
Therefore, it is easy to show that reproducing the $n$-sheet cylinder manifold is impossible. 
This result means that the gauge fields cannot build a similar relation to the geometry.
Hence it implies why the back-reaction cannot generate an $n$-sheet cylinder manifold. 
Later we will show that the $n$-sheet cylinder manifold appears from the back-reaction in the SL(2) CS formulation \cite{Ammon:2013hba,deBoer:2013vca}. 
This study demonstrates why the SL(2) CS formulation is a non-trivial holographic model.

\section{SL(2) CS Formulation}
\label{sec:3}
 \noindent
 We first introduce the action in the SL(2) CS formulation and the construction of the AdS$_3$ geometry from the gauge fields \cite{Witten:1988hc}. 
 The boundary action is a deformation of the U(1) case. 
 In the end, we introduce the Wilson line to demonstrate the building of the ``minimum surface=entanglement entropy'' \cite{Huang:2019nfm}. 
 This study shows that the back-reaction produces the $n$-sheet cylindrical manifold \cite{Ammon:2013hba}. 
 It exhibits the difference between the gauge fields and geometry in the SL(2) CS formulation and U(1) CS theory. 
 We also provide construction with two wedges to realize the Hayward term \cite{Hayward:1993my,Takayanagi:2019tvn} in the SL(2) CS formulation.
 
 \subsection{Action}
 \noindent
The action of the SL(2) CS formulation is given by \cite{Witten:1988hc}
\bea
&&S_{\mathrm{G}}
\nn\\
&=&\frac{k}{2\pi}\int d^3x\ \mathrm{Tr}\bigg(A_tF_{r\theta}-\frac{1}{2}\big(A_r\partial_tA_{\theta}-A_{\theta}\partial_tA_r\big)\bigg)
\nn\\
&&-\frac{k}{2\pi}\int d^3x\ \mathrm{Tr}\bigg(\bar{A}_t\bar{F}_{r\theta}-\frac{1}{2}\big(\bar{A}_r\partial_t\bar{A}_{\theta}-\bar{A}_{\theta}\partial_t\bar{A}_r\big)\bigg)
\nn\\
&&-\frac{k}{4\pi}\int dtd\theta\ \mathrm{Tr}\bigg(\frac{E_t^+}{E_{\theta}^+}A_{\theta}^2\bigg)
\nn\\
&&+\frac{k}{4\pi}\int dtd\theta\ \mathrm{Tr}\bigg(\frac{E_t^-}{E_{\theta}^-}\bar{A}_{\theta}^2\bigg), 
\eea
where 
\bea
F_{\mu\nu}^a\equiv\partial_{\mu}A_{\nu}^c-\partial_{\nu}A_{\mu}^a+\lbrack A_{\mu}, A_{\nu}\rbrack^a. 
\eea 
Another SL(2) field strength $\bar{F}_{\mu\nu}$ has a similar definition by replacing $A$ with $\bar{A}$. 
We label the index of Lie algebra by $c$.  
The boundary conditions of the gauge fields are defined by:
\bea
(E_{\theta}^+A_t-E_t^+A_{\theta})|_{r\rightarrow\infty}=0; \qquad (E_{\theta}^-\bar{A}_t-E_t^-\bar{A}_{\theta})|_{r\rightarrow\infty}=0.
\eea
To identify the SL(2) CS formulation by the 3d pure Einstein gravity theory, one defines the constant $k$ as
\bea
k\equiv\frac{l}{4G_3},
\eea
where 
\bea
\frac{1}{l^2}\equiv-\Lambda.
\eea
Note that the $\Lambda$ is the cosmological constant, and the $G_3$ is the three-dimensional gravitational constant. The $F_{r\theta}$ and $\bar{F}_{r\theta}$ are the $r$-$\theta$ components of the field strengths associated with the gauge potential $A$ and $\bar{A}$, respectively. 
The gauge fields are relevant to the vielbein $e_{\mu}$ and spin connection $\omega_{\mu}$, but the gauge group is now SL(2):
\bea
A_{\mu}\equiv A_{\mu}^aJ_a\equiv \frac{1}{l}e_{\mu}+\omega_{\mu}; \qquad 
\bar{A}_{\nu}\equiv \bar{A}_{\nu}^a\bar{J}_a\equiv\frac{1}{l} e_{\nu}-\omega_{\nu}.
\eea
The indices of bulk spacetime and Lie algebra are raised or lowered by 
\bea
\eta\equiv\mathrm{diag}(-1,1,1).
\eea 
The bulk theory is (locally) equivalent to the standard CS theory with the SL(2) gauge group. 
The measure in this gauge formulation is $\int {\cal D}A{\cal D}\bar{A}$, not the same as in the metric formulation.
\\

\noindent
We now summarize our conventions here. 
The SL(2)$\times$SL(2) generators are given by:
\bea
&&J_0\equiv\begin{pmatrix}
0&-\frac{1}{2}\\
\frac{1}{2}&0
\end{pmatrix}; \qquad 
J_1\equiv\begin{pmatrix}
0&\frac{1}{2}\\
\frac{1}{2}&0
\end{pmatrix}; 
\nn\\
&&J_2\equiv\begin{pmatrix}
\frac{1}{2}&0\\
0&-\frac{1}{2}
\end{pmatrix},
\nn\\
&&\bar{J}_0\equiv\begin{pmatrix}
0&-\frac{1}{2}\\
\frac{1}{2}&0
\end{pmatrix}; \qquad 
\bar{J}_1\equiv\begin{pmatrix}
0&-\frac{1}{2}\\
-\frac{1}{2}&0
\end{pmatrix}; 
\nn\\
&&\bar{J}_2\equiv\begin{pmatrix}
\frac{1}{2}&0\\
0&-\frac{1}{2}
\end{pmatrix}.
\eea
The generators satisfy the below algebraic relations: 
\bea
&&\lbrack J^a, J^b\rbrack=\epsilon^{abc}J_c; \qquad \mathrm{Tr}\big(J^aJ^b\big)=\frac{1}{2}\eta^{ab}, 
\nn\\
&&\lbrack \bar{J}^a, \bar{J}^b\rbrack=-\epsilon^{abc}\bar{J}_c; \qquad \mathrm{Tr}\big(\bar{J}^a \bar{J}^b\big)=\frac{1}{2}\eta^{ab}.
\eea
\\
 
\noindent
We will work within the AdS$_3$ background, and the geometry is given by
\bea
ds^2=-(r^2+1)dt^2+\frac{dr^2}{r^2+1}+r^2d\theta^2,
\eea
in which the ranges of coordinates are defined by:
\bea
-\infty< t<\infty; \qquad 0<r<\infty; \qquad 0<\theta\le 2\pi.
\eea
Once we choose the unit 
 \bea
 \Lambda=-1,
 \eea 
 the AdS$_3$ solution corresponds to the following gauge fields:
\bea
A&=&\sqrt{r^2+1}J_0dx^++rJ_1dx^++\frac{dr}{\sqrt{r^2+1}}J_2;
\nn\\
\bar{A}&=&\sqrt{r^2+1}\bar{J}_0dx^-+r\bar{J}_1dx^-+\frac{dr}{\sqrt{r^2+1}}\bar{J}_2.
\eea
The configuration of gauge fields satisfies the equations of motion: 
\bea
F=\bar{F}=0,
\eea 
and the solutions can be represented by the SL(2) transformation, $g$ and $\bar{g}$.

 \subsection{Boundary Theory}
 \noindent
 We first show that the 2d Schwarzian theory appears on the boundary \cite{Cotler:2018zff}, and the boundary theory is dual to another theory, which extends the U(1) case by the interaction.
 
 \subsubsection{2d Schwarzian Theory}
  \noindent
  To obtain the boundary effective theory, we first integrate out $A_0$ and $\bar{A}_0$, which is equivalent to using the following condition, respectively:
  \bea
  F_{r\theta}=0; \qquad \bar{F}_{r\theta}=0.
  \eea 
  We substitute the condition into the action and consider AdS$_3$ geometry with a boundary. 
  One can obtain the boundary manifold from the asymptotic behavior of the gauge fields:
  \bea
A_{r\rightarrow\infty}=
\begin{pmatrix}
\frac{dr}{2r}& 0
\\
rE^+& -\frac{dr}{2r}
\end{pmatrix};
\qquad
\bar{A}_{r\rightarrow\infty}=
\begin{pmatrix}
-\frac{dr}{2r}& -rE^{-}
\\
0& \frac{dr}{2r}
\end{pmatrix},
\nn\\
\eea
We show the boundary manifold of AdS$_3$ geometry by the $E^+$ and $E^-$.
The parameterization of the SL(2) transformations is the same as the following:
\bea
g_{\mathrm{SL(2)}}&=&
\begin{pmatrix}
1& 0
\\
F& 1
\end{pmatrix}
\begin{pmatrix}
\lambda & 0
\\
0& \frac{1}{\lambda}
\end{pmatrix}
\begin{pmatrix}
1 &\Psi
\\
0& 1
\end{pmatrix}; 
\nn\\
\bar{g}_{\mathrm{SL(2)}}&=&
\begin{pmatrix}
1& -\bar{F}
\\
0& 1
\end{pmatrix}
\begin{pmatrix}
\frac{1}{\bar{\lambda}} & 0
\\
0& \bar{\lambda}
\end{pmatrix}
\begin{pmatrix}
1 &0
\\
-\bar{\Psi}& 1
\end{pmatrix}.
\eea
The asymptotic gauge fields provide the constraint on the boundary from the identification:
\bea
g^{-1}_{\mathrm{SL(2)}}\partial_{\theta}g_{\mathrm{SL(2)}}|_{r\rightarrow\infty}
=A_{\theta}|_{r\rightarrow\infty}, \qquad 
\bar{g}^{-1}_{\mathrm{SL(2)}}\partial_{\theta}\bar{g}_{\mathrm{SL(2)}}|_{r\rightarrow\infty}
=\bar{A}_{\theta}|_{r\rightarrow\infty}.
\eea
This identification leads to the following constraint on the boundary:
\bea
\lambda=\sqrt{\frac{r E_{\theta}^+}{\partial_{\theta}F}}; \ 
\Psi=-\frac{1}{2rE_{\theta}^+}\frac{\partial_{\theta}^2F}{\partial_{\theta}F}, \qquad 
\bar{\lambda}=\sqrt{\frac{r E_{\theta}^-}{\partial_{\theta}\bar{F}}}; \  
\bar{\Psi}=-\frac{1}{2rE_{\theta}^-}\frac{\partial_{\theta}^2\bar{F}}{\partial_{\theta}\bar{F}}. 
\eea
The boundary action is
 \bea
&&S_{\mathrm{Gb}}
\nn\\
&=&\frac{k}{2\pi}\int dtd\theta\ \bigg(\frac{3}{2}\frac{(D_-\partial_{\theta}{\cal F})(\partial_{\theta}^2{\cal F})}{(\partial_{\theta}{\cal F})^2}-\frac{D_-\partial_{\theta}^2{\cal F}}{\partial_{\theta}{\cal F}}
\bigg)
\nn\\
&&-\frac{k}{2\pi}\int dtd\theta\ \bigg(\frac{3}{2}\frac{(D_+\partial_{\theta}\bar{{\cal F}})(\partial_{\theta}^2\bar{{\cal F}})}{(\partial_{\theta}\bar{{\cal F}})^2}-\frac{D_+\partial_{\theta}^2\bar{{\cal F}}}{\partial_{\theta}\bar{{\cal F}}}
\bigg),
\eea
where we have defined:
\bea
{\cal F}\equiv\frac{F}{E_{\theta}^+}; \qquad \bar{{\cal F}}\equiv\frac{\bar{F}}{E_{\theta}^-},
\eea
 \bea
D_+\equiv\frac{1}{2} \partial_t-\frac{1}{2}\frac{E_{t}^-}{E_{\theta}^-}\partial_{\theta}; \qquad D_-\equiv\frac{1}{2}\partial_t-\frac{1}{2}\frac{E_{t}^+}{E_{\theta}^+}\partial_{\theta}.
\eea
The $E_{\theta}^{\pm}$ is a constant here. 
Our computation is only on the flat torus in this paper. 
When computing the partition function on a spherical manifold, we also use the Lagrangian from the flat torus manifold. 
We ignore a total derivative term from the Lagrangian. 
Therefore, we can relax the condition. 
The $E_{\theta}^{\pm}$ can depend on $t$. 
Note that the boundary theory loses the conformal symmetry but remains Weyl invariant (keeping ${\cal F}$ and $\bar{{\cal F}}$ invariant). 
The Weyl symmetry is necessary for mapping a single interval to a spherical manifold or $n$-sheet cylinder manifold for calculating entanglement entropy \cite{Casini:2011kv}. 
 
 \subsubsection{Dual Theory on Boundary}
 \noindent
 Now we show that the boundary theory is dual to the following action
\bea
S_{2d1}=\frac{4k}{\pi}\int dt d\theta\ \bigg(\big(D_-\phi\big)\big(\partial_{\theta}\phi\big)+\Pi\big(\partial_{\theta}{\cal F}-e^{4\phi}\big)\bigg).
\eea
The measure of the path integral is $\int d\phi d{\cal F} d\Pi.$ 
Now we first integrate out the $\Pi$ and then integrate out the $\phi$, equivalent to using the equalities:
\bea
&&\ln\partial_{\theta}{\cal F}=4\phi; \qquad 
D_{-}\phi=\frac{1}{4}\frac{D_-\partial_{\theta}{\cal F}}{\partial_{\theta}{\cal F}}; \qquad
\partial_{\theta}\phi=\frac{1}{4}\frac{\partial_{\theta}^2{\cal F}}{\partial_{\theta}{\cal F}};
\nn\\
&&
\big(D_-\phi\big)\big(\partial_{\theta}\phi\big)
=
\frac{1}{16}\frac{\partial_{\theta}^2{\cal F}}{(\partial_{\theta}{\cal F})^2}\big(D_-\partial_{\theta}{\cal F}\big),
\eea
and then we obtain 
\bea
S_{2d2}=\frac{k}{4\pi}\int dtd\theta\  \frac{\partial_{\theta}^2{\cal F}}{(\partial_{\theta}{\cal F})^2}\big(D_-\partial_{\theta}{\cal F}\big).
\eea
The measure becomes $\int (d{\cal F}/\partial_{\theta}{\cal F})$.
We can show that the dual theory is equivalent to the 2d Schwarzian theory up to a total derivative term
\bea
\frac{k}{2\pi}\int dtd\theta\ \bigg(\frac{3}{2}\frac{(D_-\partial_{\theta}{\cal F})(\partial_{\theta}^2{\cal F})}{(\partial_{\theta}{\cal F})^2}-\frac{D_-\partial_{\theta}^2{\cal F}}{\partial_{\theta}{\cal F}}\bigg)
=\frac{k}{4\pi}\int dtd\theta\ \frac{\partial_{\theta}^2{\cal F}}{(\partial_{\theta}{\cal F})^2}(D_-\partial_{\theta}{\cal F}). 
\nn\\
\eea
Now we integrate out the ${\cal F}$, and then the generic solution of $\Pi$ is 
\bea
\Pi=f(t).
\eea 
Then we can integrate out $\Pi$ and perform a field redefinition
\bea
\phi\rightarrow\phi-\frac{1}{4}\ln f
\eea
 to obtain the following action
\bea
S_{2d3}=\frac{2k}{\pi}\int dtd\theta\ \bigg(\big(\partial_t\phi\big)\big(\partial_{\theta}\phi\big)
-\frac{E_t^+}{E_{\theta}^+}\big(\partial_{\theta}\phi\big)\big(\partial_{\theta}\phi\big)
-e^{4\phi}\bigg),
\eea
up to a total derivative term. 
The measure is $\int d\phi$.
\\

\noindent
Now we do a similar dual for the $\bar{A}$ part and begin from the following action
\bea
S_{2d1t}=\frac{4k}{\pi}\int dt d\theta\ \bigg(\big(D_+\bar{\phi}\big)\big(\partial_{\theta}\bar{\phi}\big)+\bar{\Pi}\big(\partial_{\theta}\bar{{\cal F}}-e^{4\bar{\phi}}\big)\bigg).
\eea
The measure is $\int d\bar{\phi}d\bar{{\cal F}}d\bar{\Pi}$.
We first integrate out the field, $\bar{{\cal F}}$  and then the generic solution of $\bar{\Pi}$ becomes
\bea
\bar{\Pi}=-\bar{f}(t).
\eea
Then we can do a field redefinition
\bea
\bar{\phi}\rightarrow\bar{\phi}-\frac{1}{4}\ln\bar{f}
\eea
 to obtain the following action
\bea
S_{2d3t}=\frac{2k}{\pi}\int dtd\theta\ \bigg(\big(\partial_t\bar{\phi}\big)\big(\partial_{\theta}\bar{\phi}\big)
-\frac{E_t^-}{E_{\theta}^-}\big(\partial_{\theta}\bar{\phi}\big)\big(\partial_{\theta}\bar{\phi}\big)
+e^{4\bar{\phi}}\bigg).
\eea
The measure is $\int d\bar{\phi}$.
\\

\noindent
Hence the 2d Schwarzian theory is dual to the theory:
\bea
&&S_{CB}
\nn\\
&=&
S_{2d3}+S_{2d3t}
\nn\\
&=&\frac{4k}{\pi}\int dt d\theta\ \bigg(\big(D_-\phi\big)\big(\partial_{\theta}\phi\big)-e^{4\phi}\bigg)
-\frac{4k}{\pi}\int dt d\theta\ \bigg(\big(D_+\bar{\phi}\big)\big(\partial_{\theta}\bar{\phi}\big)+e^{4\bar{\phi}}\bigg).
\nn\\
\eea
The measure is $\int d\phi d\bar{\phi}$. 
Note that the U(1) case loses the interacting terms, $\exp(4\phi)$ and $\exp(4\bar{\phi})$. 
One can first do the field redefinition:
\bea
\Pi\rightarrow\frac{1}{k}\Pi; \qquad \bar{\Pi}\rightarrow\frac{1}{k}\bar{\Pi}.
\eea
One then absorbs the $k$ into $\phi$ and $\bar{\phi}$ as in the following: 
\bea
\phi\rightarrow\frac{\phi}{\sqrt{k}}; \qquad \bar{\phi}\rightarrow\frac{\bar{\phi}}{\sqrt{k}}.  
\eea
The interacting term vanishes (or does not depend on $\phi$) under the weak gravitational constant limit $k\rightarrow\infty$. 
Hence all new information should be encoded by the interacting terms. 
We will do a one-loop exact calculation for entanglement entropy \cite{Huang:2019nfm} to show the difference between the SL(2) CS formulation and U(1) CS theory.

\subsection{Wilson Line}
\noindent
We calculate the Wilson line \cite{Ammon:2013hba}
\bea
&&
W_{\cal R}(C)
\nn\\
&=&
\int DU DP D\lambda\ 
\exp\bigg\lbrack\int_C ds\ \bigg(\mathrm{Tr}(PU^{-1}D_s U)
+\lambda(s)\big(\mathrm{Tr}(P^2)-c_2\big)\bigg)\bigg\rbrack,
\nn\\
\eea
where $U$ is now an SL(2) element, $P$ is its conjugate momentum, 
\bea
\sqrt{2c_2}\equiv \frac{c}{6}(1-n),
\eea
 but $c$ becomes the central charge of CFT$_2$. 
 The covariant derivative is:
\bea
D_sU\equiv \frac{d}{ds}U+A_sU+U\bar{A}_s, \ A_s\equiv A_{\mu}\frac{dx^{\mu}}{ds}.
\eea 
The endpoints of the Wilson line are at the end of a single interval on the boundary of AdS$_3$.
The trace operation in the $W_{\cal R}(C)$ acts on the representation. 
The equations of motion are:
\bea
i\frac{k}{2\pi}F_{\mu_1\mu_2}&=&-\int ds\ \frac{dx^{\mu_3}}{ds}\epsilon_{\mu_1\mu_2\mu_3}\delta^3\big(x-x(s)\big)UPU^{-1};
\nn\\ 
i\frac{k}{2\pi}\bar{F}_{\mu_1\mu_2}&=&\int ds\ \frac{dx^{\mu_3}}{ds}\epsilon_{\mu_1\mu_2\mu_3}\delta^3\big(x-x(s)\big)P.
\label{eq:eom}
\eea
One solution of the above equations of motion is related to the SL(2) transformations, $g$ and $\bar{g}$ \cite{Ammon:2013hba}: 
\bea
&&A=g^{-1}ag+g^{-1}dg, \ g=\exp(L_1 z)\exp(\rho L_0); 
\nn\\
&&\bar{A}=-\bar{g}^{-1}a\bar{g}-\bar{g}^{-1}d\bar{g}, \ \bar{g}=\exp(L_{-1}\bar{z})\exp(-\rho L_0),
\eea
where
\bea
a\equiv\frac{1}{k}\sqrt{\frac{c_2}{2}}\bigg(\frac{dz}{z}-\frac{d\bar{z}}{\bar{z}}\bigg)L_0.
\eea
The equations of motion are similar to the U(1) case. 
The generation of the holonomy and solution has the same situation.
The SL(2) algebra satisfies the following relations: 
\bea
&&
\lbrack L_j, L_k\rbrack=(j-k)L_{j+k}, \ j, k=0,\pm 1; 
\nn\\
&&
 \mathrm{Tr}(L_0^2)=\frac{1}{2}; \qquad \mathrm{Tr}(L_{-1}L_1)=-1,
\nn\\
\eea
 and the traces of other bilinears vanish.
We use a different basis to express the SL(2) algebra because it is more convenient to show the solution. 
We then choose: 
\bea
z\equiv r\exp(i\Phi); \qquad \bar{z}\equiv r\exp(-i\Phi)
\eea
to obtain the spacetime \cite{Ammon:2013hba}
\bea
ds^2=d\rho^2+\exp({2\rho})(dr^2+n^2r^2d\Phi^2).
\eea 
This solution corresponds to the following choice: 
\bea
U(s)=1; \qquad P(s)=\sqrt{2c_2}L_0
\eea 
with the curve: 
\bea
z(s)=0; \qquad \rho(s)=s.
\eea
When $n$ approaches one, the geometry is AdS$_3$. 
The back-reaction of Wilson line generates the $n$-sheet manifold \cite{Ammon:2013hba}. 
Using the redefinition
\bea
r\equiv \exp(t),
\eea
the boundary geometry becomes the $n$-sheet cylinder
\bea
ds^2=dt^2+n^2d\Phi^2
\eea
up to a Weyl transformation. 
Because the boundary theory is scale-invariant, one can apply a scale transformation to geometry. 
\\

\noindent
Because $W_{\cal R}$ vanishes under the limit $n\rightarrow 1$, only the pure Einstein gravity theory survives. 
One can see that the Wilson line serves as a useful auxiliary tool for analytical continuation. 
The expectation value of the Wilson line is
\bea
\langle W_{\cal R}\rangle=\frac{Z_n}{Z_1^n}+{\cal O}(n-1),
\eea
 where $Z_n$ is the $n$-sheet partition function of the boundary theory. 
 When we do the analytical continuation of $n$ to one, the bulk calculation is equivalent to the boundary calculation. 
 The analytical continuation exactly shows the operator correspondence between the bulk operator, Wilson line, and entanglement entropy of the boundary theory
\bea
S_{EE}=\lim_{n\rightarrow 1}\frac{1}{1-n}\ln \langle W_{\cal R}\rangle,
\eea
Because the classical solution gives the entanglement entropy of CFT$_2$ \cite{deBoer:2013vca}, our result provides the quantum deformation of a geodesic line. 
\\

\noindent
If entanglement entropy is proportional to the $\langle\ln W_{\cal R}\rangle$, the quantum contribution only contributes to an area term. 
However, our result is instead $\ln\langle W_{\cal R}\rangle$. 
Hence holographic entanglement entropy \cite{Ryu:2006bv} cannot be given by the area term. 
One cannot observe the difference between the two expressions at the classical level. 
Therefore, the study of quantum correction is a non-trivial task. 
\\

\noindent
We discuss why the ``minimum surface=entanglement entropy'' can work in the SL(2) CS formulation. 
We first introduce the Wilson line. 
The back-reaction gives the $n$-sheet geometry. 
The back-reaction is the same as a deformation of the background solution or gauge fields as in the U(1) case. 
The geometry is relevant to the gauge fields now. 
The deformation changes the geometry, and then the $n$-sheet manifold appears. 
Because the background changes, the boundary zweibein also needs to change without changing the boundary condition. 
Deforming the boundary zweibein also changes the boundary term of the SL(2) CS formulation. 
Indeed, it is non-trivial. 
Now the back-reaction of the Wilson line leads to a natural modification. 
The boundary theory depends on the choice of background. 
We cannot obtain an exact solution (effective action) when introducing the Wilson line, but we can in the U(1) case. 
It is hard to know whether the limit $n\rightarrow 1$ is smooth without an exact solution. 
Therefore, a similar exact study should be helpful.
In the U(1) case, we show that the Wilson line deforms the chiral scalar fields. 
An additional change to the boundary condition is necessary for obtaining entanglement entropy. 

\subsection{Hayward Term}
\noindent
Here we discuss an alternative approach (the co-dimension two Hayward term) for computing entanglement entropy. 
The Hayward term is \cite{Hayward:1993my}
\bea
S_H\equiv\frac{1}{8\pi G_3}\int_{\Gamma}ds\ (\pi-\tilde{\theta}),
\eea
where $s$ is the proper length of the curve 
\bea
\Gamma\equiv\Sigma_1\cap\Sigma_2,
\eea
with co-dimension one boundary surfaces $\Sigma_1$ and $\Sigma_2$. 
 The $\tilde{\theta}$ is the interior angle between two surfaces, $\Sigma_1$ and $\Sigma_2$, is 
 \bea
-\pi<\tilde{\theta}\le\pi.
\eea 
Thus, the $\Gamma$ represents the co-dimension two wedge. 
 The Hayward term ensures a well-posed variational principle when a boundary manifold has wedges.
\\

\begin{figure}
\begin{center}
\includegraphics[width=8cm]{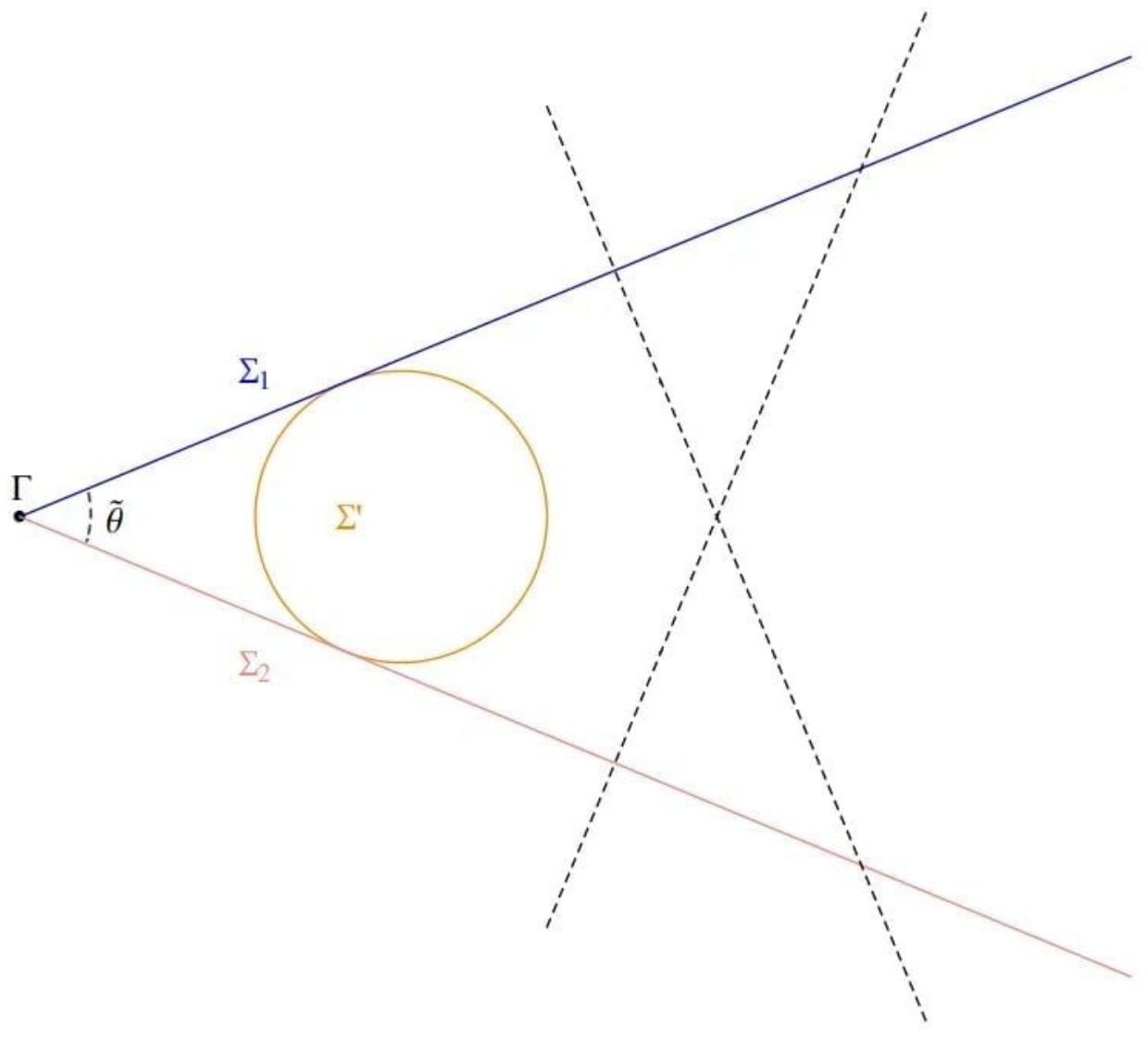}
\end{center}
\caption{The $\Sigma_1$ and $\Sigma_2$ are the co-dimension one boundary surfaces. 
The $\tilde{\theta}$ is the interior angle between two surfaces, $\Sigma_1$ and $\Sigma_2$.
The $\Gamma=\Sigma_1\cap\Sigma_2$ is the co-dimension two wedge term. 
The $\Sigma^{\prime}$ is a smooth 2d surface.
}
\label{Haywardfig.pdf}
\end{figure}

\noindent
We consider the setup in Fig. \ref{Haywardfig.pdf} and replace the joint $\Gamma$ with a cap of radius $r$ by introducing a local smooth co-dimension one surface $\Sigma^{\prime}$. 
 Note that we take 
\bea
r=\epsilon,
\eea
and eventually, we would shrink $\epsilon$ to zero. 
Now we are only considering a thin slice of the joint. 
Therefore, one can always assume that the metric is locally flat around the joint. 
Therefore, the boundary condition at $r=\epsilon$ becomes:
\bea
(A_t-A_{\theta})|_{r=\epsilon\rightarrow 0}=0; \qquad (\bar{A}_t+\bar{A}_{\theta})|_{r=\epsilon\rightarrow 0}=0. 
\label{bdh}
\eea
One can solve the wedge term by the boundary integration in the co-dimension one surface  $\Sigma^{\prime}$ under the limit 
\bea
\Sigma^{\prime} \to \Gamma
\eea
 that collapses into integration in the co-dimension two joint \cite{Hayward:1993my}.
\\

\noindent
We introduce the boundary term
\bea
S_{HCS}\equiv
-\frac{k}{4\pi}\int_{\Sigma^{\prime}} dtd\theta\ \mathrm{Tr}(A_{\theta}^2)
-\frac{k}{4\pi}\int_{\Sigma^{\prime}} dtd\theta\ \mathrm{Tr}(\bar{A}_{\theta}^2)
\label{bdy}
\eea
 at
\bea
r=\epsilon,
\eea
instead of at infinity. 
The solutions of the gauge fields at the boundary are:
\bea
A=J_0dx^++J_2dr; \qquad 
\bar{A}=\bar{J}_0dx^-+\bar{J}_2dr.
\eea
These solutions provide the following relation:
\bea
&&
g^{-1}\partial_{\theta}g|_{r=0}
\nn\\
&=&
\begin{pmatrix}
\frac{1}{\lambda}\partial_{\theta}\lambda-\Psi\lambda^2\partial_{\theta}F &\partial_{\theta}\Psi+2\frac{\Psi}{\lambda}\partial_{\theta}\lambda-\Psi^2\lambda^2\partial_{\theta}F
\\
\lambda^2\partial_{\theta}F&\lambda^2\Psi\partial_{\theta}F-\frac{1}{\lambda}\partial_{\theta}\lambda
\end{pmatrix}\bigg|_{r=0}
\nn\\
&=&
A_{\theta}|_{r=0}
\nn\\
&=&
\begin{pmatrix}
0& -\frac{1}{2}
\\
\frac{1}{2}&0
\end{pmatrix}.
\eea
This equality shows the following conditions:
\bea
\frac{1}{\lambda}\partial_{\theta}\lambda&=&\Psi\lambda^2\partial_{\theta}F; 
\nn\\
\lambda^2\partial_{\theta}F&=&\frac{1}{2};
\nn\\
\partial_{\theta}\Psi+2\frac{\Psi}{\lambda}\partial_{\theta}\lambda-\Psi^2\lambda^2\partial_{\theta}F&=&-\frac{1}{2}.
\eea
Hence the conditions imply the following equations:
\bea
\lambda^2=\frac{1}{2\partial_{\theta}F}; \qquad \frac{2}{\lambda}\partial_{\theta}\lambda=\Psi; \qquad
\partial_{\theta}\Psi+\frac{1}{2}\Psi^2=-\frac{1}{2}.
\eea
After we substitute the boundary conditions to the bulk theory with the boundary term, we obtain the 2d Schwarzian theory for the $F$. 
For the $\bar{F}$, the result is similar.
Hence combining the bulk theory and the boundary term gives
\bea
&&S_{BH}
\nn\\
&=&\frac{k}{2\pi}\int dtd\theta\ \bigg(\frac{3}{2}\frac{\partial_-\partial_{\theta}F\partial_{\theta}^2F}{(\partial_{\theta}F)^2}
-\frac{\partial_-\partial_{\theta}^2F}{\partial_{\theta}F}\bigg)
\nn\\
&&
-\frac{k}{2\pi}\int dtd\theta\ \bigg(\frac{3}{2}\frac{\partial_+\partial_{\theta}\bar{F}\partial_{\theta}^2\bar{F}}{(\partial_{\theta}\bar{F})^2}
-\frac{\partial_+\partial_{\theta}^2\bar{F}}{\partial_{\theta}\bar{F}}\bigg).
\nn\\
\eea
However, substituting the classical solution to the bulk term and Hayward term shows that only the boundary term survives. 
In the end, we obtain:
\bea
S_{BH}=-\frac{k}{2\pi}\int dtd\theta\ \mathrm{Tr}(e_{\theta}^2+\omega_{\theta}^2)=\frac{1}{2} S_H.
\eea
The boundary term does not lead to the correct wedge term \cite{Rooman:2000zi}. 
One can trace it back to the fact that the usual boundary term \eqref{bdy}, when recast in the gauge formulation, is already off by a factor of two. 
\\

\begin{figure}
\begin{center}
\includegraphics[width=8cm, height=8cm]{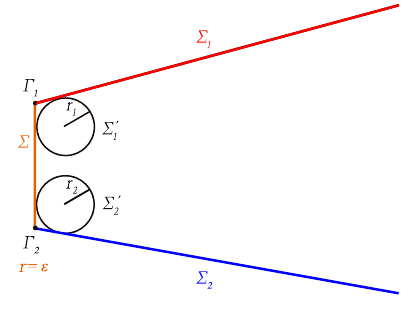}
\end{center}
\caption{A construction with double wedges. 
The $\Sigma_1$ and $\Sigma_2$ are the co-dimension one boundary surfaces. 
The $\Sigma$ is a cut-off surface at $r=\epsilon$. 
We replace the two joints $\Gamma_1$ and $\Gamma_2$ with two caps of radius $r_1$ and $r_2$, respectively.}
\label{Haywardfig2.pdf}
\end{figure}

\noindent
A way to fix this problem at the classical level is to consider double wedges. 
The setup is in Fig. \ref{Haywardfig2.pdf}. 
In this scenario, we impose co-dimension one surfaces $\Sigma_1$ and $\Sigma_2$ with the boundary condition \eqref{bdh}. 
However, an additional cut-off surface is at 
\bea
r=\epsilon.
\eea
 We immediately see that this would introduce two cusp-like co-dimension two wedges. 
 Following a similar procedure, we replace the two joints $\Gamma_1$ and $\Gamma_2$ with the two caps of the radius $r_1$ and $r_2$ respectively. 
 We take two limits in the following order. 
 We first shrink the two caps to zero radius to obtain the correct Hayward term. 
 Eventually, we take the cut-off surface 
 \bea
 \epsilon \to 0.
 \eea 
 This limit provides a natural interpretation for considering a classical gravity dual of a CFT on a single interval. 
The two approaches differ in the order of the limits.
\\

\begin{figure}
\begin{center}
\includegraphics[width=15cm, height=15cm]{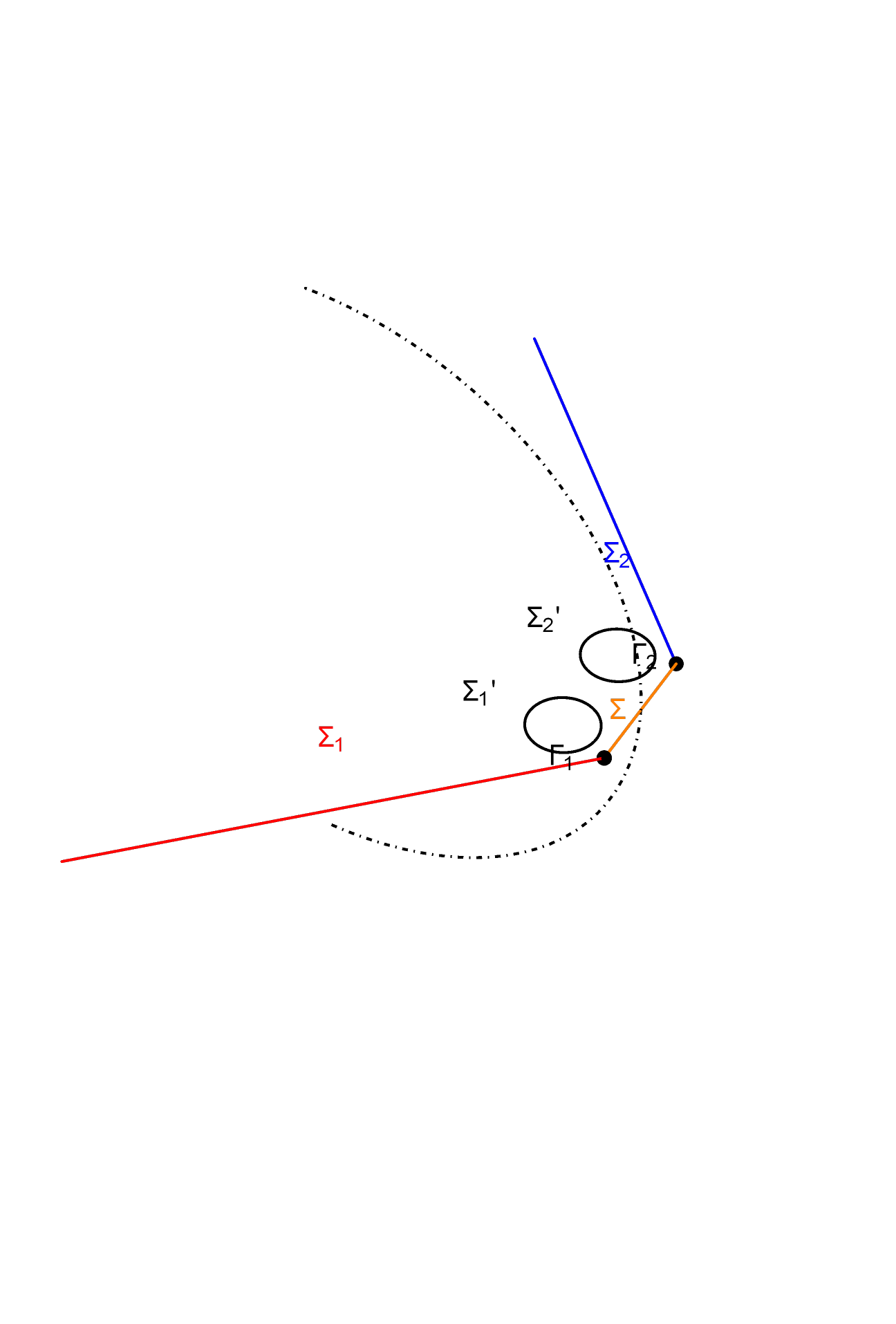}
\end{center}
\caption{
The dotted line is the AdS$_3$ minimum surface. 
The wedges are $\Gamma_1$ and $\Gamma_2$. 
When $r_1, r_2=\epsilon\rightarrow 0$, two wedges approach to the minimum surface. 
The $\Sigma_1^{\prime}$ and $\Sigma_2^{\prime}$ are the co-dimension one cut-off surfaces. 
The $\Sigma$ is a cut-off surface at $r=\epsilon$. 
}
\label{Hayward3D.pdf}
\end{figure}

\noindent
Classically, we have recovered the Hayward term. 
A natural question is whether one can use the wedge picture to construct the $n$-sheet manifold and perform a well-defined analytic continuation of $n \to 1$ in the quantum regime. 
We solve the fluctuation field in the equations of motion. 
Note that the equations of motion of the fluctuation field $F$ are:
\bea
\lambda^2=\frac{1}{2\partial_{\theta}F}; \qquad \frac{2}{\lambda}\partial_{\theta}\lambda=\Psi; \qquad
\partial_{\theta}\Psi+\frac{1}{2}\Psi^2=-\frac{1}{2}.
\eea
We can rewrite the equations as the following
\bea
\partial^2_\theta (\ln \partial_\theta F)-\frac{1}{2}(\partial_\theta (\ln \partial_\theta F))^2=\frac{1}{2}.
\eea
It is easy to obtain the general solution by:
\bea
f \equiv \ln \partial_\theta F=-2 \ln \cos{\bigg( \frac{\theta}{2}+c_1(t) \bigg) }+c_2(t), 
\eea
where $c_1(t)$ and $c_2(t)$ are arbitrary time-dependent functions.
By doing the integration for $\theta$, we obtain the solution of the $F(\theta)$ as
\be
F(\theta)=2 \exp\big(c_2(t)\big) \tan{\bigg( \frac{\theta}{2}+c_1(t) \bigg)}+c_3(t),
\label{Fsol}
\ee
where $c_3(t)$ is also an arbitrary time-dependent function.
The solution is periodic for $\theta$
\bea
F(\theta) = F(\theta+2 \pi).
\eea 
When $\epsilon\rightarrow 0$, the wedge sits at the location of the AdS$_3$ minimum surface. 
In Fig. \ref{Hayward3D.pdf}, we show the location of wedges compared to the minimum surface. 
Therefore, we can find how the minimum surface is generated from the wedges. 
The wedge method can provide boundary entanglement entropy.  
We will discuss the result of the quantum contribution in the next section.

\section{Entanglement Entropy at One-Loop}
\label{sec:4}
\noindent
We calculate entanglement entropy for a single interval (with the length $L$) in the boundary theory of the SL(2) CS formulation \cite{Huang:2019nfm}. 
For the convenience of calculation, we choose another form of the boundary theory on a spherical manifold \cite{Cotler:2018zff}. 
Note that the calculation on the torus manifold is one-loop exact \cite{Cotler:2018zff}. 
Therefore, the entanglement entropy is also one-loop exact.
The entanglement entropy shows a shift of central charge, 26 \cite{Huang:2019nfm}, and the shifting does not appear in the U(1) case. 
In the end, we discuss the quantum correction for the Hayward approach \cite{Hayward:1993my}. 
The result shows that the quantum correction vanishes. 
Therefore, we have seen the difference between the approaches of the Wilson line \cite{Ammon:2013hba} and Hayward terms \cite{Takayanagi:2019tvn} in the quantum regime.  
The details of Rényi entropy for the one-loop correction are in Appendix \ref{appa}.

\subsection{Boundary Theory}
\noindent
After we integrate out $A_t$ and $\bar{A}_t$, the theory shows the form of the SL(2) transformations
\bea
&&S_{\mathrm{G}1}
\nn\\
&=&-\frac{k}{4\pi}
\nn\\
&&\times\int d^3x\ \mathrm{Tr}\bigg(-g^{-1}\big(\partial_rg\big)g^{-1}\big(\partial_tg\big)g^{-1}\partial_{\theta}g
+g^{-1}\big(\partial_rg\big)g^{-1}\big(\partial_{\theta}g\big)g^{-1}\big(\partial_tg\big)
\bigg)
\nn\\
&&+\frac{k}{4\pi}
\nn\\
&&\times\int d^3x\ \mathrm{Tr}\bigg(-\bar{g}^{-1}\big(\partial_r\bar{g}\big)\bar{g}^{-1}\big(\partial_t\bar{g}\big)\bar{g}^{-1}
\partial_{\theta}\bar{g}
+\bar{g}^{-1}\big(\partial_r\bar{g}\big)\bar{g}^{-1}\big(\partial_{\theta}\bar{g}\big)\bar{g}^{-1}\big(\partial_t\bar{g}\big)
\bigg)
\nn\\
&&+\frac{k}{2\pi}\int dt d\theta\ \mathrm{Tr}\bigg(g^{-1}\big(\partial_{\theta}g\big)g^{-1}\big(D_-g\big)\bigg)
\nn\\
&&-\frac{k}{2\pi}\int dt d\theta\ \mathrm{Tr}\bigg(\bar{g}^{-1}\big(\partial_{\theta}\bar{g}\big)\bar{g}^{-1}\big(D_+\bar{g}\big)\bigg).
\eea
By rewriting the SL(2) transformations in terms of other parameters, the effective action becomes:
\bea
&&S_{\mathrm{G}1}
\nn\\
&=&\frac{k}{\pi}\int dtd\theta\ \bigg(\frac{(\partial_{\theta}\lambda)(D_-\lambda)}{\lambda^2}+\lambda^2(\partial_{\theta}F)(D_-\Psi)\bigg)
\nn\\
&&
-\frac{k}{\pi}\int dtd\theta\ \bigg(\frac{(\partial_{\theta}\bar{\lambda})(D_+\bar{\lambda})}{\bar{\lambda}^2}+\bar{\lambda}^2(\partial_{\theta}\bar{F})(D_+\bar{\Psi})\bigg).
\eea
Due to the following boundary conditions: 
\bea
\lambda^2\partial_{\theta}F=E_{\theta}^+r; \qquad \bar{\lambda}^2\partial_{\theta}\bar{F}=E_{\theta}^-r,
\eea
the following terms: 
\bea
\frac{k}{\pi}\int dtd\theta\ \bigg(\lambda^2(\partial_{\theta}F)(D_-\Psi)\bigg), \qquad
-\frac{k}{\pi}\int dtd\theta\ \bigg(\bar{\lambda}^2(\partial_{\theta}\bar{F})(D_+\bar{\Psi})\bigg)
\eea
are total derivative terms.
Therefore, the boundary effective action becomes
\bea
S_{\mathrm{G}1}
=\frac{k}{\pi}\int dtd\theta\ \bigg(\frac{(\partial_{\theta}\lambda)(D_-\lambda)}{\lambda^2}
-\frac{(\partial_{\theta}\bar{\lambda})(D_+\bar{\lambda})}{\bar{\lambda}^2}\bigg).
\eea
\\

\noindent
With the definition: 
\bea
{\cal F}\equiv\frac{F}{E^+_{\theta}}; \qquad
\bar{{\cal F}}\equiv\frac{\bar{F}}{E_{\theta}^-},
\eea
we obtain:
\bea
\lambda=\sqrt{\frac{r}{\partial_{\theta}{\cal F}}}; \qquad
\bar{\lambda}=\sqrt{\frac{r}{\partial_{\theta}\bar{{\cal F}}}}.
\eea
Therefore, we obtain the boundary action
\bea
S_{\mathrm{G}1}
=\frac{k}{4\pi}\int dtd\theta\ \bigg(\frac{(\partial_{\theta}^2{\cal F})(D_-\partial_{\theta}{\cal{F}})}{(\partial_{\theta}{\cal F})^2}
-\frac{(\partial_{\theta}^2\bar{{\cal F}})(D_+\partial_{\theta}\bar{{\cal{F}}})}{(\partial_{\theta}\bar{{\cal F}})^2}\bigg).
\eea
Finally, we choose the field redefinition:
\bea
{\cal F}\equiv \tan\bigg(\frac{\phi}{2}\bigg); \qquad \bar{{\cal F}}\equiv\tan\bigg(\frac{\bar{\phi}}{2}\bigg)
\label{fredef}
\eea
to obtain
\bea
&&S_{\mathrm{G}1}
\nn\\
&=&\frac{k}{4\pi}\int dtd\theta\ \bigg\lbrack\frac{(\partial_{\theta}^2\phi)(D_-\partial_{\theta}\phi)}{(\partial_{\theta}\phi)^2}
-(\partial_{\theta}\phi)(D_-\phi)\bigg\rbrack
\nn\\
&&
-\frac{k}{4\pi}\int dtd\theta\ \bigg\lbrack\frac{(\partial_{\theta}^2\bar{\phi})(D_+\partial_{\theta}\bar{\phi})}{(\partial_{\theta}\bar{\phi})^2}
-(\partial_{\theta}\bar{\phi})(D_+\bar{\phi})\bigg\rbrack,
\eea
up to a total derivative term.

\subsection{Spherical Manifold}
\noindent
Using Weyl transformations, the Euclidean AdS$_3$ metric approaching to a boundary ($r\rightarrow\infty$) can be written as \cite{Casini:2011kv}
\bea
ds^2=\frac{dr^2}{r^2}+r^2ds_s^2,
\eea
where 
\bea
ds_s^2\equiv d\psi^2+\sin^2\psi d\theta^2, \ 0\le\psi<\pi, \ 0\le\theta<2\pi.
\eea
The boundary zweibein is:
\bea
E_{\theta}^+=E_{\theta}^-=\sin\psi; \qquad E_{\psi}^+=-iE_t^+=i; \qquad E_{\psi}^-=-iE_t^-=-i.
\eea

\subsection{Entanglement Entropy for Single Interval}
\noindent
To calculate entanglement entropy for a single interval, we first identify the boundary conditions of fields. 
We then calculate entanglement entropy and compare the result to the U(1) case. 
We calculate the R\'enyi entropy 
\begin{equation}
\label{eq:REE}
S_n\equiv\frac{\ln Z_n-n\ln Z_1}{1-n}
\end{equation}
 from the replica trick \cite{Holzhey:1994we},
where $Z_n$ is the $n$-sheet partition function, and $Z_1$ is equivalent to the partition function. 
When $n\rightarrow 1$, the R\'enyi entropy gives entanglement entropy. 
The calculation is one-loop exact. 
Therefore, the quantum fluctuation of the $n$-sheet partition function only comes from the one-loop order. 
We then obtain that the $n$-sheet partition function is a product of the classical $n$-sheet partition-function ($Z_{n, c}$) and the one-loop $n$-sheet partition-function ($Z_{n, q}$) 
\bea
Z_n=Z_{n, c}\cdot Z_{n, q}.
\eea 
Calculating the R\'enyi entropy is necessary to take the logarithm on the $n$-sheet partition function
\bea
\ln Z_n=\ln Z_{n, c}+\ln Z_{n, q}.
\eea
Hence the contribution from the classical and the one-loop terms is not mixed.

\subsubsection{$n$-Sheet Manifold}
\noindent
We first do a coordinate transformation on the unit sphere $ds_s^2$ to obtain 
\bea
ds_s^2=\sech^2(y)(dy^2+d\theta^2),
\eea 
where
\bea
\sech y\equiv\sin\psi.
\eea
The range of $\theta$ in the $n$-sheet manifold becomes
\bea
0<\theta\le 2\pi n.
\eea 
The periodicity of $\theta$  becomes $2\pi n$ now. 
The $y$-direction needs regularization, and the range becomes
\bea
-\ln\frac{L}{\epsilon}< y\le \ln\frac{L}{\epsilon}, 
\eea 
where $\epsilon$ is a cut-off. 
We sum two momentum modes with different boundary conditions. 
One mode satisfies the Dirichlet boundary condition in the $y$-direction. 
Another one follows the Neumann boundary conditions.   
Therefore, the periodicity is $4\ln(L/\epsilon)$.

\subsubsection{Boundary Condition}
\noindent
Now we map the sphere to the torus. 
We choose the coordinates of an $n$-sheet torus 
\bea
z\equiv\frac{\theta+iy}{n}
\eea
 with the identification: 
\bea
z\sim z+2\pi; \qquad z\sim z+2\pi\tau_{n}.
\eea
The identification leads to the boundary condition \cite{Huang:2019nfm}: 
\bea
\phi\bigg(\frac{y}{n}, \frac{\theta}{n}+2\pi \bigg)&=&\phi\bigg(\frac{y}{n}, \frac{\theta}{n}\bigg)+2\pi;  
\nn\\
\phi\bigg(\frac{y}{n}+2\pi\cdot\mathrm{Im}(\tau_{n}), \frac{\theta}{n}+2\pi\cdot\mathrm{Re}(\tau_{n})\bigg)&=&
\phi\bigg(\frac{y}{n}, \frac{\theta}{n}\bigg), 
\nn\\ 
\bar{\phi}\bigg(\frac{y}{n}, \frac{\theta}{n}+2\pi\bigg)&=&\bar{\phi}\bigg(\frac{y}{n}, \frac{\theta}{n}\bigg)+2\pi; 
\nn\\
\bar{\phi}\bigg(\frac{y}{n}+2\pi\cdot\mathrm{Im}(\tau_{n}), \frac{\theta}{n}+2\pi\cdot\mathrm{Re}(\tau_{n})\bigg)&=&
\bar{\phi}\bigg(\frac{y}{n}, \frac{\theta}{n}\bigg).
 \eea
Due to the periodicity of $y$, the complex structure corresponds to the unit sphere is \cite{Huang:2019nfm}
\bea
\tau_{n}=\bigg(\frac{2i}{n\pi}\bigg)\ln\bigg(\frac{L}{\epsilon}\bigg).
\eea
Note that 
\bea
\mathrm{Re}(\tau_n)=0.
\eea
The Fourier transformation of the fields gives: 
\bea
\phi=\frac{\theta}{n}+\epsilon(y, \theta); \qquad \bar{\phi}=-\frac{\theta}{n}+\bar{\epsilon}(y, \theta),
\eea
where
\bea
\epsilon(y, \theta)&\equiv&\sum_{j, k}\epsilon_{j, k} e^{i\frac{j}{n}\theta-\frac{k}{\tau}y}; \qquad
\epsilon_{j, k}^*\equiv\epsilon_{-j, -k},
\nn\\
\bar{\epsilon}(y, \theta)&\equiv&\sum_{j, k}\bar{\epsilon}_{j, k} e^{i\frac{j}{n}\theta-\frac{k}{\tau}y}; \qquad
\bar{\epsilon}_{j, k}^*\equiv\bar{\epsilon}_{-j, -k}.
\eea
The saddle-point is the $\theta/n$ in the $\phi$ and the $-\theta/n$ in the $\bar{\phi}$.
Each Fourier mode has three zero-modes: 
\bea
\epsilon_{j, k}=0; \qquad \bar{\epsilon}_{j, k}=0, \ j=-1, 0, 1
\eea 
because of the SL(2) gauge symmetry.

\subsubsection{Entanglement Entropy}
\noindent
To calculate the $n$-sheet partition function, we need to Wick rotate for the boundary effective action:
\bea
&&S_{\mathrm{EG}1}
\nn\\
&=&-\frac{k}{4\pi}\int_0^{\pi} d\psi\int _0^{2\pi n} d\theta\ \bigg\lbrack\frac{(\partial_{\theta}^2\phi)(D_-\partial_{\theta}\phi)}{(\partial_{\theta}\phi)^2}
-(\partial_{\theta}\phi)(D_-\phi)\bigg\rbrack
\nn\\
&&
+\frac{k}{4\pi}\int_0^{\pi} d\psi\int_0^{2\pi n} d\theta\ \bigg\lbrack\frac{(\partial_{\theta}^2\bar{\phi})(D_+\partial_{\theta}\bar{\phi})}{(\partial_{\theta}\bar{\phi})^2}
-(\partial_{\theta}\bar{\phi})(D_+\bar{\phi})\bigg\rbrack
\nn\\
&=&\frac{k}{4\pi}\int_{-\frac{\pi}{2}\mathrm{Im}(\tau)}^{\frac{\pi}{2}\mathrm{Im}(\tau)} dy \int_0^{2\pi n} d\theta\ \sech(y)
\bigg\lbrack\frac{(\partial_{\theta}^2\phi)(D_-\partial_{\theta}\phi)}{(\partial_{\theta}\phi)^2}
-(\partial_{\theta}\phi)(D_-\phi)\bigg\rbrack
\nn\\
&&
-\frac{k}{4\pi}\int_{-\frac{\pi}{2}\mathrm{Im}(\tau)}^{\frac{\pi}{2}\mathrm{Im}(\tau)} dy \int _0^{2\pi n}d\theta\ \sech(y)
 \bigg\lbrack\frac{(\partial_{\theta}^2\bar{\phi})(D_+\partial_{\theta}\bar{\phi})}{(\partial_{\theta}\bar{\phi})^2}
-(\partial_{\theta}\bar{\phi})(D_+\bar{\phi})\bigg\rbrack,
\nn\\
\eea
in which the covariant derivative becomes:
\bea
D_+&=&\frac{i}{2}\partial_{\psi}-\frac{i}{2}\frac{E_{\psi}^-}{E_{\theta}^-}\partial_{\theta}
=\frac{i}{2}\partial_{\psi}-\frac{1}{2\sin\psi}\partial_{\theta}
=-\frac{i}{2}
\cosh(y)\partial_y-\frac{1}{2}\cosh(y)\partial_{\theta}, 
\nn\\
D_-&=&\frac{i}{2}\partial_{\psi}-\frac{i}{2}\frac{E_{\psi}^+}{E_{\theta}^+}\partial_{\theta}
=\frac{i}{2}\partial_{\psi}+\frac{1}{2\sin\psi}\partial_{\theta}
=-\frac{i}{2}
\cosh(y)\partial_y+\frac{1}{2}\cosh(y)\partial_{\theta}.
\nn\\
\eea
\\

\noindent
We substitute the saddle-point into the boundary effective action and note that 
\bea
c=6k
\eea
as the central charge of CFT$_2$, we have:
\bea
&&-\frac{k}{4\pi}\int_{-\frac{\pi}{2}\mathrm{Im}(\tau)}^{\frac{\pi}{2}\mathrm{Im}(\tau)} dy \int_0^{2\pi n} d\theta\ \frac{1}{2n^2}
-\frac{k}{4\pi}\int_{-\frac{\pi}{2}\mathrm{Im}(\tau)}^{\frac{\pi}{2}\mathrm{Im}(\tau)}dy\int_0^{2\pi n}d\theta\ \frac{1}{2n^2}
\nn\\
&=&\frac{-c\pi\tau}{12n}
\nn\\
&=&-\frac{c}{6n}\ln\frac{L}{\epsilon},
\eea
where 
\bea
\tau\equiv n\tau_n.
\eea 
Therefore, we obtain
\bea
\ln Z_{n, c}=\frac{c}{6n}\ln\frac{L}{\epsilon}.
\eea
The Rényi entropy from the saddle-point is:
\bea
S_{n, c}=\frac{c}{1-n}\bigg(\frac{1}{6n}-\frac{n}{6}\bigg)\ln\frac{L}{\epsilon}
=\frac{c(1+n)}{6n}\ln\frac{L}{\epsilon}.
\eea
When we take $n\rightarrow 1$, the saddle-point contributes to the entanglement entropy
\bea
S_{1, c}=\frac{c}{3}\ln\frac{L}{\epsilon}.
\eea

\noindent
The Rényi entropy from the one-loop correction is given by:
\bea
S_{n, q}=\frac{1}{1-n}\big(\ln Z_{n, q}-n\ln Z_{1, q}\big)=\frac{13(n+1)}{3n}\ln\frac{L}{\epsilon}.
\eea
Therefore, the summation of classical and one-loop terms gives the Rényi entropy
\bea
S_{n}=\frac{(c+26)(n+1)}{6n}\ln\frac{L}{\epsilon}
\eea
and the entanglement entropy
\bea
S_{1}=\frac{c+26}{3}\ln\frac{L}{\epsilon}.
\eea
The details of the one-loop correction are in Appendix \ref{appa}.
\\

\noindent
The result shows that the quantum contribution does not change the form of Rényi entropy and entanglement entropy. 
It shifts the value of the central charge by 26. Hence the conformal anomaly does not take the Rényi entropy and entanglement entropy to go beyond the results of CFT$_2$. 
For the U(1) case, entanglement entropy does not have such a shift from the $k^0$ term, and it is proportional to $k$.
The difference between the U(1) and SL(2) cases is self-interaction. 
Therefore, the interaction vanishes in the weak gravitational constant limit $k\rightarrow\infty$.
Hence the shift of the central charge or the one-loop contribution is due to the self-interaction. 
This result shows that the SL(2) CS formulation should match the U(1) CS theory in the weak gravitational constant limit.
\\

\noindent
The U(1) theory only has two degrees of freedom, but the SL(2) formulation has  six. 
It interprets why the geometry can be related to the SL(2) formulation but not the U(1) theory. 
However, their equivalence is on the boundary. 
The bulk does not have dynamical degrees of freedom. 
The asymptotic gauge fields showing the AdS$_3$ metric give the additional four required constraints. 
Its boundary theory provides consistent degrees with the U(1) case. 
The self-interaction term indicates the difference between the U(1) and SL(2) cases. 
However, it does not affect the physical degrees of freedom. 
Hence studying the exact solution in the U(1) case provides a better understanding of the SL(2) case.

\subsection{Hayward Term}
\noindent
We discuss whether it is possible to study the quantum correction by computing the bulk and Hayward terms. 
From the solution \eqref{Fsol}, we choose the saddle point as in the Wilson line:
\bea
F_{saddle}\equiv 2\exp\big(c_2(t)\big)\tan\bigg(\frac{\theta}{2}\bigg)+c_3(t).
\eea 
By substituting the solution into the action with the field redefinition \eqref{fredef}, the $c_2(t)$ and $c_3(t)$ would not affect the result. 
Because the introduction of the $n$-sheet torus with the boundary conditions for the $\phi$ would be the same, the quantum fluctuation only comes from the $c_1(t)$ term in Eq. \eqref{Fsol}. 
All situations are similar to the Wilson line, except that the quantum fluctuation does not depend on the $\theta$. 
The non-vanishing one-loop result relies on the non-trivial dependence of the $\theta$. 
Therefore, the partition function does not include quantum fluctuation. 
 Introducing the $n$-sheet manifold to the approach of the Hayward term \cite{Takayanagi:2019tvn}, the quantum fluctuation still cannot generate the dependence of the $\theta$. 
 Therefore, the result implies the inconsistency between the entanglement entropy and naive quantum generalization to the quantum regime. 
 This fact shows why the classical term is too universal, and including quantum contribution is a non-trivial task.   

\section{Discussion and Conclusion}
\label{sec:5}
\noindent
In this paper, we studied holographic entanglement entropy \cite{Ryu:2006bv}. 
We considered the SL(2) CS formulation \cite{Witten:1988hc} for the AdS$_3$ Einstein gravity theory and compared it to the U(1) CS theory. 
We then showed that the Wilson line is a suitable bulk operator for obtaining entanglement entropy of the boundary theory \cite{Huang:2019nfm,Ammon:2013hba,deBoer:2013vca}. 
This proof is quite non-trivial because it is non-perturbative. 
We also showed that the Hayward term \cite{Hayward:1993my} in the SL(2) CS formulation reproduces the entanglement entropy at the classical level by a double-wedges construction. 
Combining the bulk and Hayward terms for a quantum generalization provides the vanishing quantum correction in the partition function.  
Hence the Wilson line is a suitable candidate for studying entanglement entropy.
The proof strengthens the correspondence of ``minimum surface=entanglement entropy'', but it relies on a smooth limit of the analytical continuation \cite{Huang:2019nfm}. 
Hence it is not clear why such equivalence can appear. 
This result motivates us to find a simple system to do a similar exact study. 
We can study the Wilson line in the U(1) CS theory to establish the equivalence. 
The equivalence is necessary to deform the boundary after the back-reaction of the Wilson line. 
The chiral scalar fields are not relevant to geometry (in the U(1) CS theory). 
Therefore, the deformation of boundary conditions and back-reaction cannot appear simultaneously. 
However, it does not imply that the boundary fields can reconstruct kinematic information (boundary geometry) in the U(1) CS theory. 
It is mainly because the numbers of gauge fields on the bulk are not equivalent. 
The weak gravitational constant limit does not play a role in defining geometry.
In the SL(2) CS formulation, the back-reaction deforms the gauge fields and also the geometry \cite{Ammon:2013hba,deBoer:2013vca}. 
Because the boundary zweibein is coupled to the boundary gauge fields, deforming the geometry is equivalent to changing the boundary condition and theory. 
We calculated entanglement entropy in the SL(2) CS formulation. 
The result showed a shift of the central charge by 26 \cite{Huang:2019nfm}. 
Therefore, the shift should be due to the self-interaction term (because the U(1) case loses it).
\\

\noindent
The U(1) CS theory has two independent gauge fields, but the SL(2) CS formulation has six gauge fields. 
It seems that their physical degrees of freedom do not match. 
However, the bulk theory in a topological theory does not have dynamics. 
We need to compare the physical degrees of freedom in the boundary theories. 
In the SL(2) CS formulation, the AdS$_3$ geometry constrains the asymptotic gauge fields. 
The constraint leads the boundary theory to have the equivalent physical degrees of freedom to the U(1) CS theory. 
The SL(2) CS formulation has one self-interacting term in the boundary theory. 
The term does not appear in the U(1) case. 
However, the local interacting term does not change the physical degrees of freedom. 
Because the interacting term vanishes in the weak gravitational constant limit, the pure AdS$_3$ Einstein gravity theory approaches the U(1) CS theory. 
The reduction itself is not trivial. 
The various exact results also complement the SL(2) CS formulation. 
\\

\noindent
We comment on the holographic principle in the SL(2) CS formulation. 
For obtaining entanglement entropy, the bulk calculation reduces to the boundary calculation. 
The study provides a better understanding of the holographic principle. 
The classical contribution is usually not enough to justify different proposals or conjectures. 
The holographic principle works here because it requires one necessary condition, the coupling between boundary fields and the asymptotic metric field. 
By studying the U(1) CS theory, we know that the condition is not trivial. 
In this case, changing the boundary condition (or Lagrangian) after introducing the geodesic operator (Wilson lines) is necessary. 
\\ 

\noindent 
In the end, we discuss other local AdS spaces. 
In general other locally AdS spaces like BTZ black hole can be obtained from pure AdS by a quotient of some discrete subgroup $\Gamma$ of the isometry group PSL$(2,\mathbb{C})$. 
The correspondence between pure Einstein gravity and the Schwarzian theory on the boundary remains valid in any event. 
In this sense, we expect that our computation of quantum correction can be generalized to other backgrounds like AdS black holes. 
We have Riemann surfaces of higher genus after the replica trick. 
Explicit computation by solving eigenvalues is beyond the scope of the current manuscript. 
One can obtain the form of the one-loop correction from earlier results. 
It is well-known that the one-loop contribution from the Schwarzian theory agrees with the bulk one-loop partition function of graviton \cite{Cotler:2018zff}. 
In a general background $\mathbb{H}^3/\Gamma$, the bulk one-loop partition function can be obtained as a product form \cite{Giombi:2008vd}
\bea
Z^{\mathbb{H}^3/\Gamma} = \prod_{\gamma \in {\cal P}} Z^{\mathbb{H}^3/\langle \gamma\rangle},\label{one-loopgl}
\eea
where the product is over the primitive conjugate classes of $\Gamma$. 
Each $Z^{\mathbb{H}^3/\langle \gamma\rangle}$ is the one-loop partition on a solid torus and hence agrees with our one-loop contribution (it follows from the boundary Schwarzian theory on a torus). 
It is natural to expect that the one-loop contribution from the boundary Schwarzian theory always agrees with that of the bulk graviton.  Hence the quantum correction given by Eq. \eqref{one-loopgl} on the right-hand side is explicitly known. 
 
\section*{Acknowledgments}
\noindent
We thank Chuan-Tsung Chan, Bartlomiej Czech, Jan de Boer, Kristan Jensen, and Ryo Suzuki for their discussion. 
Chen-Te Ma would thank Nan-Peng Ma for his encouragement.
\\

\noindent
Xing Huang acknowledges the support of the NSFC Grants No. 11947301 and No. 12047502. 
Chen-Te Ma acknowledges the YST Program of the APCTP;  
Post-Doctoral International Exchange Program; 
China Postdoctoral Science Foundation, Postdoctoral General Funding: Second Class (Grant No. 2019M652926); 
Foreign Young Talents Program (Grant No. QN20200230017).
Hongfei Shu acknowledges the JSPS Research Fellowship 17J07135 for Young Scientists from the Japan Society for the Promotion of Science (JSPS); the grant “Exact Resultsin Gauge and String Theories” from the Knut and Alice Wallenberg foundation. 
Chih-Hung Wu acknowledges the National Science Foundation under Grant No. 1820908 and the Ministry of Education, Taiwan (R. O. C).
\\

\noindent
 We thank the National Tsing Hua University, Institute for Advanced Study at the Tsinghua University, and the Center for Quantum Science at the Sogang University.
 \\
 
 \noindent
Discussion during the workshops, ``East Asia Joint Workshop on Fields and Strings 2019'' and ``The 17th Italian-Korean Symposium for Relativistic Astrophysics'', was helpful to this work. 

\appendix
\section{Details of Rényi Entropy for One-Loop}
\label{appa}
\noindent
Now we consider the quantum fluctuation from the $\epsilon(y, \theta)$ and $\bar{\epsilon}(y, \theta)$ 
to obtain the one-loop term \cite{Huang:2019nfm}. 
Because the contributions from the sector of $\phi$ and $\bar{\phi}$ are the same, we only show the calculation related to the $\phi$ field.
The expansion from the $\epsilon$ in the boundary effective action is \cite{Huang:2019nfm}
\bea
&&\frac{k}{4\pi}\int_{-\frac{\pi}{2}\mathrm{Im}(\tau)}^{\frac{\pi}{2}\mathrm{Im}(\tau)} dy\int_0^{2\pi n}d\theta\ \bigg(n^2\big(\partial_{\theta}^2\epsilon(y, \theta)\big)\big(\bar{\partial}\partial_{\theta}\epsilon(y, \theta)\big)
\nn\\
&&
-\big(\partial_{\theta}\epsilon(y, \theta)\big)\big(\bar{\partial}\epsilon(y, \theta)\big)\bigg)
\nn\\
&=&
-i\frac{k}{4\pi}n\tau\cdot\bigg\{ n^2\sum_{j, k}\bigg\lbrack\bigg(-\frac{j^2}{n^2}\cdot\frac{1}{2}\cdot\bigg(i\frac{k}{\tau}+i\frac{j}{n}\bigg)\bigg(i\frac{j}{n}\bigg)\bigg\rbrack|\epsilon_{j, k}|^2
\nn\\
&&
-\sum_{j ,k}\bigg\lbrack\bigg(i\frac{j}{n}\bigg)\cdot\frac{1}{2}\cdot\bigg(i\frac{k}{\tau}+i\frac{j}{n}\bigg)\bigg\rbrack|\epsilon_{j, k}|^2\bigg\}
\nn\\
&=&-i\frac{k}{8\pi}\sum_{j, k}j(j^2-1)\bigg(k+\frac{j}{n}\tau\bigg)|\epsilon_{j, k}|^2,
\eea
where
\bea
\bar{\partial}\equiv\frac{1}{2}(-i\partial_y+\partial_{\theta}).
\eea
Doing the derivative on the logarithm of the $n$-sheet one-loop partition shows 
\bea
\partial_{\tau}\ln Z_{n, q}=-\sum_{j\neq 0,\pm 1}\sum_{k=-\infty}^{\infty}\frac{\frac{j}{n}}{k+\frac{j}{n}\tau}.
\eea
Applying the following useful relation of the digamma function, we obtain
\bea
\tilde{\psi}(1-x)-\tilde{\psi}(x)=\pi\cot(\pi x),
\eea
in which the digamma function is defined by
\bea
\tilde{\psi}(a)\equiv
-\sum_{n=0}^{\infty}\frac{1}{n+a}.
\eea
Therefore, the complicated summation in the $n$-sheet partition function simplified as:
\bea
\sum_{m=-\infty}^{\infty}\frac{1}{m-x}&=&-\sum_{m=0}^{\infty}\frac{1}{m+x}+\sum_{m=1}^{\infty}\frac{1}{m-x}
\nn\\
&=&-\sum_{m=0}^{\infty}\frac{1}{m+x}+\sum_{m=0}^{\infty}\frac{1}{m+1-x}
=\tilde{\psi}(x)-\tilde{\psi}(1-x)
\nn\\
&=&-\pi\cdot\cot(\pi x).
\eea
Hence we obtain:
\bea
\partial_{\tau}\ln Z_{n, q}&=&-\sum_{j\neq 0,\pm 1}\sum_{k=-\infty}^{\infty}\frac{\frac{j}{n}}{k+\frac{j}{n}\tau}
=-\sum_{j\neq 0, \pm 1}\bigg(\frac{j}{n}\pi\bigg)\cdot\cot\bigg(\frac{j}{n}\pi\tau\bigg)
\nn\\
&=&-2\pi\sum_{j=2}^{\infty}\bigg(\frac{j}{n}\bigg)\cdot\cot\bigg(\frac{j\pi\tau}{n}\bigg).
\eea
To obtain a universal term, we use the re-summation:
\bea
\partial_{\tau}\ln Z_{n, q}
&=&-2\pi\sum_{j=2}^{\infty}\bigg(\frac{j}{n}\bigg)\cdot\cot\bigg(\frac{j\pi\tau}{n}\bigg)
\nn\\
&=&-2\pi\sum_{j=2}^{\infty}\frac{j}{n}\cdot\bigg\lbrack\cot\bigg(\frac{j\pi\tau}{n}\bigg)+i\bigg\rbrack
+2\pi i\sum_{j=2}^{\infty}\frac{j}{n}.
\nn\\
\eea
We can perform a regularization for the divergent series
\bea
\sum_{j=1}^{\infty}j\rightarrow -\frac{1}{12}.
\eea
Hence we obtain: 
\bea
\partial_{\tau}\ln Z_{n, q}&=&-2\pi\sum_{j=2}^{\infty}\frac{j}{n}\cdot\bigg\lbrack\cot\bigg(\frac{j\pi\tau}{n}\bigg)+i\bigg\rbrack
+2\pi i\sum_{j=2}^{\infty}\frac{j}{n}
\nn\\
&\rightarrow&-2\pi\sum_{j=2}^{\infty}\frac{j}{n}\cdot\bigg\lbrack\cot\bigg(\frac{j\pi\tau}{n}\bigg)+i\bigg\rbrack
-i\frac{13\pi }{6n}.
\eea
\\

\noindent
After integrating out the $\tau$, we obtain
\bea
\ln Z_{n, q}=-2\sum_{j=2}^{\infty}\bigg\lbrack\ln\sin\bigg(\frac{\pi j\tau}{n}\bigg)+i\frac{j\pi\tau}{n}\bigg\rbrack-i\frac{13\pi\tau}{6n}+\cdots,
\eea
where $\cdots$ is independent of the $\tau$. The first series is convergent for the 
\bea
\mathrm{Im}(\tau)>0.
\eea
When we consider the limit 
\bea
\frac{L}{\epsilon}\rightarrow\infty,
\eea
 we obtain
\bea
\ln Z_{n, q}=\frac{13}{3n}\ln\frac{L}{\epsilon},
\eea
and the Rényi entropy for the one-loop correction is \cite{Huang:2019nfm}:
\bea
S_{n, q}=\frac{1}{1-n}\big(\ln Z_{n, q}-n\ln Z_{1, q}\big)=\frac{13(n+1)}{3n}\ln\frac{L}{\epsilon}.
\eea

\end{document}